\documentclass[twocolumn,preprintnumbers,superscriptaddress,amsmath,amssymb]{revtex4}
\usepackage{graphicx}
\usepackage{dcolumn}
\usepackage{bm}
\usepackage{amssymb}
\usepackage{amssymb}
\usepackage{verbatim}
\usepackage{color}
\def\>{\rangle}

\begin{document}

\title{Optoacoustic entanglement in a continuous Brillouin-active solid state system}
\author{Changlong Zhu}
\affiliation{Max Planck Institute for the Science of Light, Staudtstra{\ss}e 2,
D-91058 Erlangen, Germany}
\author{Claudiu Genes}
\affiliation{Max Planck Institute for the Science of Light, Staudtstra{\ss}e 2,
D-91058 Erlangen, Germany}
\affiliation{Department of Physics, Friedrich-Alexander-Universit\"{a}t Erlangen-N{\"u}rnberg, Staudtstra{\ss}e 7,
D-91058 Erlangen, Germany}
\author{Birgit Stiller}\email{birgit.stiller@mpl.mpg.de}
\affiliation{Max Planck Institute for the Science of Light, Staudtstra{\ss}e 2,
D-91058 Erlangen, Germany}
\affiliation{Department of Physics, Friedrich-Alexander-Universit\"{a}t Erlangen-N{\"u}rnberg, Staudtstra{\ss}e 7,
D-91058 Erlangen, Germany}
\date{\today}

\begin{abstract}
Entanglement in hybrid quantum systems comprised of fundamentally different degrees of freedom, such as light and mechanics is of interest for a wide range of applications in quantum technologies. Here, we propose to engineer bipartite entanglement between traveling acoustic phonons in a Brillouin active solid state system and the accompanying light wave. The effect is achieved by applying
optical pump pulses to state-of-the-art waveguides, exciting a
Brillouin Stokes process. This pulsed approach, in a system operating in a regime orthogonal to standard optomechanical setups, allows for the generation of entangled photon-phonon pairs, resilient to thermal fluctuations. We propose an experimental platform where readout of the optoacoustics entanglement is done by the simultaneous detection of Stokes and Anti-Stokes photons in a two-pump configuration. The proposed mechanism presents an important feature in that it does not require initial preparation of the quantum ground state of the phonon mode.

\end{abstract}


\maketitle
Entanglement between two quantum systems describes a many body quantum state which is not separable and can exhibit quantum correlations even at
arbitrarily large distances~\cite{einstein1935can}. It is therefore a resource not only for a plethora of emerging quantum technologies~\cite{bouwmeester2000thephysics,braunstein2005quantum,horodecki2009quantum}
such as quantum cryptography~\cite{ekert1991quantum,jennewein2000quantum}, quantum teleportation~\cite{Bennett,Hofer},
and quantum computation~\cite{Gottesman}, but also offers a tool to deeper study and understand the classical-to-quantum boundary~\cite{Leggett}. As many experimental endeavors are easier performed under ambient conditions, it is essential to study possibilities of generating entangled quantum states robust to thermal noise even at room temperature. Cavity optomechanics~\cite{Aspelmeyer},
which provides a platform to explore quantum effects via coupling photons and phonons
at the macroscale, has been a hot topic of investigations in terms of generation
and measurement of entanglement both in theory~\cite{Vitali,Paternostro,Genes,Sarma}
and experiments~\cite{Palomaki,Ockeloen-Korppi,Marinkovic,Barzanjeh}.

Continuum optomechanical systems~\cite{Laer,Rakich}, such as Brillouin-active optical
waveguides~\cite{Wolff}, are a more recent alternative to standard optomechanical cavities, offering a viable platform for the interface between optical photons and continuously accessible groups of
acoustic phonons. Such platforms exhibit a powerful performance in quantum information processing with unprecedented acoustic and optical bandwidth compared to
cavity optomechanical systems. Especially due to the recent development in
nanofabrication, a new breed of chip-based Brillouin-active waveguides with
short length ($\rm{\sim cm}$) have been achieved experimentally~\cite{Eggleton},
which enables coherent
information transduction~\cite{Heedeuk}, information storage~\cite{Merklein1,Birgit-Stiller} and phonon cooling~\cite{Yin-Chung,Otterstrom}. However, the next logical step after the experimental demonstration of Brillouin cooling
~\cite{Otterstrom,Laura,Johnson} in continuum optomechanics is the observation of quantum optoacoustical entanglement.
\begin{figure}[t]
\centerline{
\includegraphics[width=8.6 cm, clip]{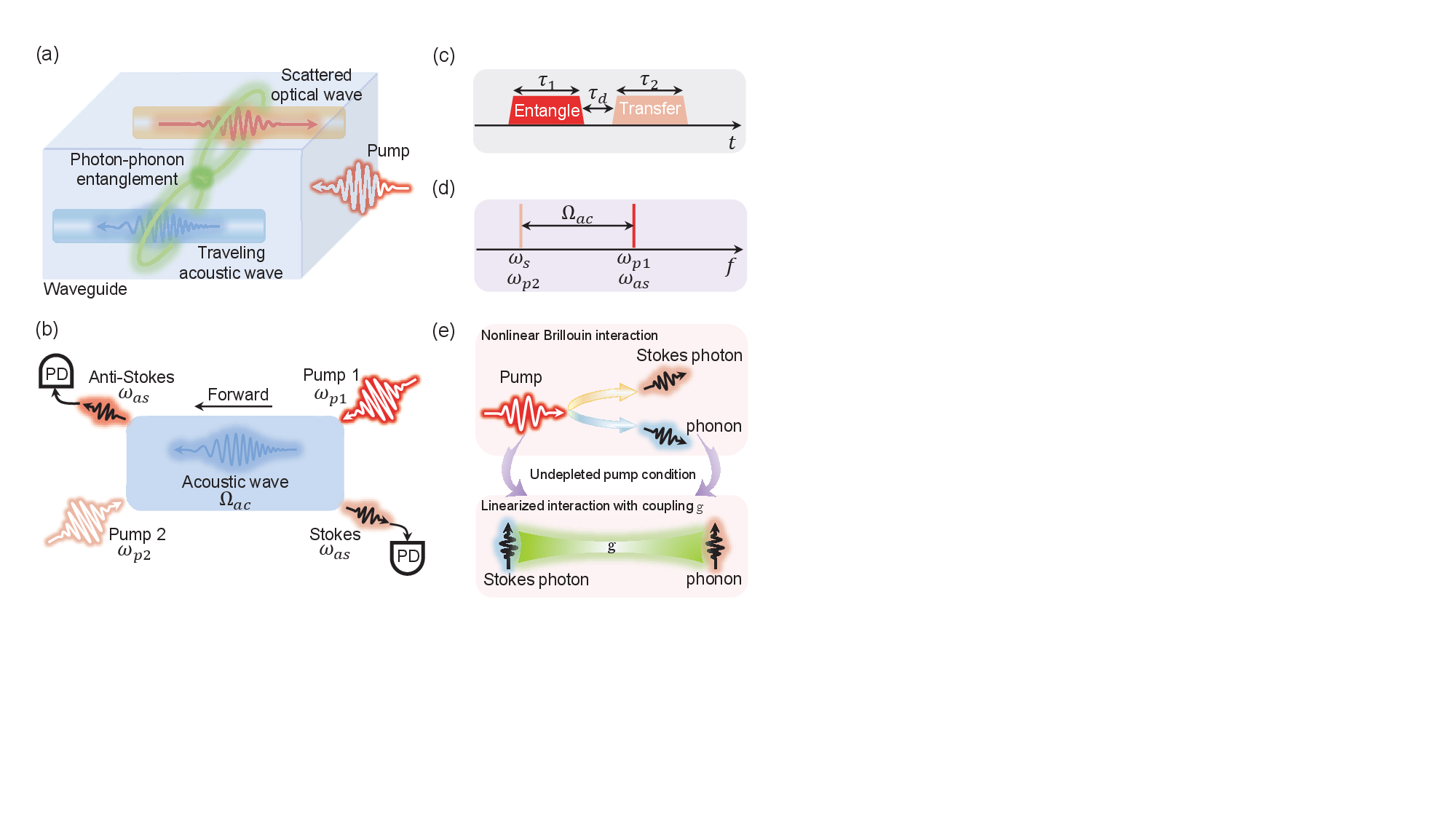}}
\caption{(a) The Brillouin Stokes
scattering implies the down conversion of a pump photon (at $\omega_\text{p}$) into a Stokes scattered photon ($\omega_\text{s}$) and traveling acoustic phonon (at $\Omega_\text{ac}$).
(b) Protocol for entanglement generation, via pump 1 and detection (at photon-diode detectors PD) via the simultaneous monitoring of both Stokes and Anti-Stokes scattered photons in the presence of pump 2.
(c) The timing diagram of the protocol showing the application of an entangling pulse for duration $\tau_1$ followed by the transfer pulse with duration $\tau_2$ after the interval $\tau_\text{d}$.
(d) The frequency relations between optical and acoustic
waves, i.e., $\omega_\text{p1}-\omega_\text{s}=\omega_\text{as}-\omega_\text{p2}=\Omega_\text{ac}$
and $\omega_\text{as}={\omega_\text{p1}}$.
(e) In the lineared regime, under the undepleted pump assumption, the nonlinear Brillouin Stokes scattering process can be mapped into a linear Hamiltonian with pump-enhanced coupling strength $g$.
}\label{Fig1}
\end{figure}

In this work, we propose and analyze the feasibility of an experimental scheme (see Fig.\ref{Fig1}) which enables the generation of bipartite light-matter entanglement in  Brillouin-active waveguides. The mechanism is based on a down-conversion like process, where pairs of phonons (frequency $\Omega_\text{ac}$) and Brillouin Stokes scattered photons (frequency $\omega_\text{s}$) are generated from the higher energy pump (with $\omega_\text{p}=\Omega_\text{ac}+\omega_\text{s}$). As opposed to standard optomechanical systems, the acoustic phonons are highly energetic (in the GHz regime) and mechanical losses (rate $\Gamma$) dominate over optical losses (rate $\gamma$). This indicates a departure from the standard continuous operation regime and suggests a pulsed operation scheme, thus circumventing the usual prerequisite of initial quantum ground state cooling~\cite{Palomaki,Ockeloen-Korppi}.
Based on experimental parameters in accordance with state-of-the-art
waveguides~\cite{Eggleton} utilized for experimental proofs of Brillouin cooling
~\cite{Otterstrom,Laura,Johnson}, we present numerical evidence of optoacoustic entanglement in the pulsed regime achieved without the need of employing non-classical quantum states of light~\cite{Pavel}. At an ambient phonon occupancy $n_\text{th}$, for small times $t<1/(\Gamma n_\text{th})$ an optimized value for the logarithmic negativity~\cite{Zyczkowski,Vidal,Adesso} is obtained in a very simple analytical form reading
$E_{\mathcal{N}}^{\rm max}\approx -{\rm ln}[1-2g^2/(\Gamma n_\text{th})^2]$.
This entanglement can be detected by homodyne~\cite{Vitali} or heterodyne detection~\cite{Palomaki} of both Stokes photons and anti-Stokes photons produced by a second pump~\cite{Junyin} (as illustrated in Fig.~\ref{Fig1}). More specifically, following the application of a forward-propagating
pump with pulse duration $\tau_1$ to stimulate the Brillouin Stokes process, a delayed (delay time $\tau_d$)
backward-propagating pulsed pump with duration $\tau_2$ is applied to stimulate the Brillouin anti-Stokes process
and thus convert the state of the acoustic phonons to a second optical pulse of
the anti-Stokes output wave. Our analytical results and numerical investigations demonstrate that this entanglement generation can be fully optically monitored with reasonable values at temperatures above the restrictive cryogenic regime.\\

\noindent \textbf{Brillouin Stokes scattering in waveguides}.--- Let us first briefly introduce the dynamics of a Brillouin Stokes
process under the condition of undepleted constant CW pump laser (pump 1) and with the observation that for backward scattering the process is independent of the anti-Stokes process because of the dispersive symmetry breaking~\cite{Laura,Kharel}.
In the undepleted pump regime, the interaction Hamiltonian can
be reduced to a linearized optoacoustic exchange between scattered photons
and acoustic phonons with an effective pump-enhanced coupling strength $g$~\cite{Laer,Yin-Chung} (illustrated in Fig.~\ref{Fig1}(e)). Envelope bosonic operators $a_\text{s}$ and $b_\text{ac}$ can be constructed,
corresponding to Stokes and acoustic waves~\cite{Kharel,Sipe,Hashem}, i.e., $a_\text{s}=1/\sqrt{2\pi}\int dk\, a(k,t) e^{-ikz}$
and $b_\text{ac}=1/\sqrt{2\pi}\int dk\, b(k,t) e^{ikz}$, where $a(k,t)$ and $b(k,t)$
denote annihilation operators for the $k$-th Stokes photon mode and acoustic
phonon mode, respectively. This formulation captures the time evolution of the amplitude
for each mode. Moving into the momentum space, the dynamics of the linearized
Stokes process can be given by~\cite{Changlong}
\begin{eqnarray}\label{Motion equation in momentum space}
\frac{d a}{d t}&=& - \left( \frac{\gamma}{2} + i \Delta_a \right)a - i g b^{\dagger} + \sqrt{\gamma}\xi_a, \nonumber\\
\frac{d b}{d t}&=& - \left( \frac{\Gamma}{2} + i \Delta_b \right)b - i g a^{\dagger} + \sqrt{\Gamma}\xi_b,
\end{eqnarray}
where $\Delta_a=k v_\text{opt}$ ($\Delta_b=k v_\text{ac}$)
denote the damping rate and wavenumber-induced frequency shift of the Stokes (acoustic) mode,
respectively. Here, $v_\text{opt}$ ($v_\text{ac}$) is the group velocity of the Stokes (acoustic)
wave. We take the effective coupling strength $g$ real and positive without loss of
generality~\cite{Laer}. The quantum noise operators $\xi_a$ and $\xi_b$ are assumed to be zero-averaged~\cite{Aspelmeyer,Yin-Chung,Kharel}
and to satisfy the following correlation $\langle \xi_{a}(t)\xi_{a}^{\dagger}(t')\rangle=\delta(t-t')$ (for the optical mode an effective zero temperature bath can be assumed) and
$\langle \xi_{b}^{\dagger}(t)\xi_{b}(t')\rangle=n_\text{th}\delta(t-t')$, where
the thermal phonon occupancy at ambient temperature $T_\text{m}$ is given by $n_\text{th}=(e^{\hbar \Omega_\text{ac}/k_\text{B}T_\text{m}} - 1)^{-1}$ ($k_\text{B}$ is the Boltzmann constant).\\

\noindent \textbf{Optoacoustic entanglement via Brillouin Stokes interaction }---In order to quantify entanglement we make use of the logarithmic negativity for continuous variables $E_{\mathcal{N}}$ introduced in Refs.~\cite{Zyczkowski,Vidal,Adesso} and extensively used in Refs.~\cite{Vitali,Rogers,Wang,Tian,Shasha,Fengxiao}.
To this end, we introduce position and momentum quadratures
$x_a=(a+a^\dagger)/\sqrt{2}$, $p_a=i(a^\dagger-a)/\sqrt{2}$, $x_b=(b+b^\dagger)/\sqrt{2}$,
and $p_b=i(b^\dagger-b)/\sqrt{2}$. The covariance matrix is then computed $\mathcal{V}_{ij}=( \langle \phi_i(t)\phi_j(t) + \phi_j(t)\phi_i(t) \rangle )/2 -
\langle \phi_i(t) \rangle \langle \phi_j(t) \rangle$, where the indexes $i$ and $j$ go over the vector $\phi^T(t)=( x_a(t),p_a(t),x_b(t),p_b(t))$ which
is the vector of continuous variables operators at time $t$. The symmetric matrix $\mathcal{V}$
can be expressed as the $2\times2$ block form
\begin{equation}
\mathcal{V} = \left[
\begin{array}{cc}
A & C\\
C^T & B
\end{array}
\right].
\end{equation}
From this, one computes
$E_{\mathcal{N}} = {\rm max}\left[ 0, -{\rm ln} (2\lambda_{-}) \right]$
where $\lambda_{-}$ is the minimal symplectic eigenvalue of the covariance matrix $\mathcal{V}$
under a partial transposition~\cite{Pirandola} and defined as $\lambda_{-}\equiv 2^{-1/2}
\sqrt{ \Sigma(\mathcal{V}) - [ \Sigma(\mathcal{V})^2 - 4\det\mathcal{V} ]^{1/2} }$,
with $\Sigma(\mathcal{V}) \equiv \det A + \det B -2\det C$.
As a general criterion for bimodal Gaussian states, entanglement needs the conditions $E_{\mathcal{N}}>0$,
which is equivalent to $\lambda_{-}<1/2$. Notice that in the particular case of a two mode squeezed state, the logarithmic negativity is simply proportional to the squeezing parameter~\cite{Zippilli,Zippilli2}.\\
\begin{figure*}[t]
  \centering
  \includegraphics[width=1.00\textwidth]{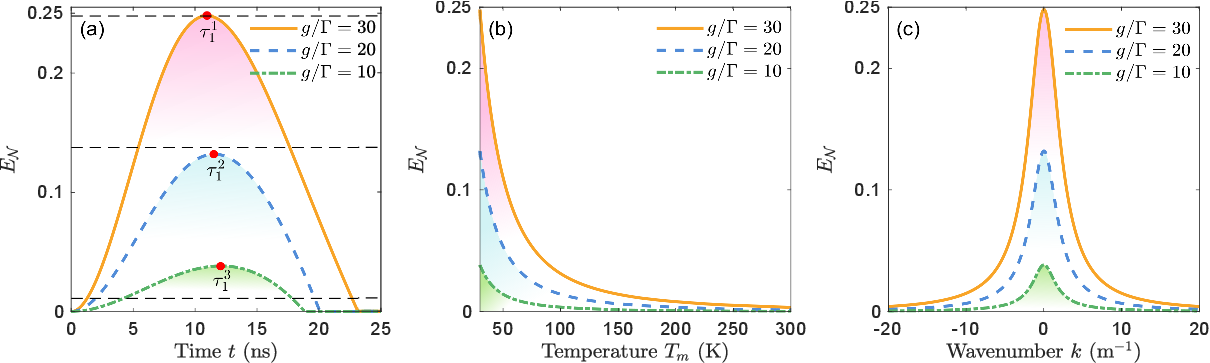}
  \caption{(a) Time evolution of $E_{\mathcal{N}}$
at $T_\text{m}=30~$K for various ratios $g/\Gamma$. The red points
denote the optimal time for optimal entanglement and black dashed
lines correspond to the maximum value of $E_{\mathcal{N}}$ evaluated
by the analytical expression.
(b) Variation of $E_{\mathcal{N}}$
versus $T_\text{m}$ for various ratios $g/\Gamma$.
(c) Continuum optoacoustic entanglement versus the wavenumber $k$ in the strong coupling
regime at temperature of $30~$K.}
\label{Fig2}
\end{figure*}
Numerically and analytically one can start with Eqs.~\eqref{Motion equation in momentum space} to compute the time evolution of the covariance matrix. We perform numerical simulations with experimentally feasible values of system parameters~\cite{Eggleton,Laura,Merklein}:
$\Gamma/2\pi=2~{\rm MHz}$, $\gamma/2\pi=0.1~{\rm MHz}$, $\Omega_\text{ac}/2\pi=7.7~{\rm GHz}$,
$c=3\times10^{8}~{\rm m/s}$, $n_\text{opt}=2.4$ (index of refraction), $v_\text{ac}=6000~{\rm m/s}$, $T_\text{m}=30~{\rm K}$,
and $\Delta_a=0.2\Gamma$. The results are illustrated in Fig.~\ref{Fig2}(a) for various ratios of $g/\Gamma$ deep into the strong coupling regime and indicate that a given time window is available for efficient optoacoustic entanglement generation.
We assume that the wavenumber-induced frequency shifts of Stokes and acoustic modes
are within the linewidth of the acoustic mode ($\Delta_{a,b}<\Gamma$), where
$\Delta_a\gg\Delta_b$ since $v_\text{opt}=c/n_\text{opt}\gg v_\text{ac}$.
The Stokes mode is assumed to be initially in
the vacuum state while a thermal state with phonon occupation $n_0=n_\text{th}\gg1$ is assumed for the acoustic mode. Under such conditions, a crude approximation for the minimal symplectic eigenvalue $\lambda_{-}$ at a high
environment temperature is given by the following expression
\begin{eqnarray}\label{Approximated_solution_of_lambda}
\lambda_{-} \approx \frac{1}{2}\left[1-2g^2t^2+\frac{2}{3}g^2(\Gamma n_\text{th})t^3\right].
\end{eqnarray}
A better analytical fit to the exact behavior can be obtained, as listed in the Appendix, albeit in a quite cumbersome form. However, the simplified expression above suffices to understand the mechanisms leading to the generation and suppression of entanglement. While the initial beam splitter exchange can lead to the generation of entanglement by reducing the symplectic eigenvalue below $1/2$, at a rate indicated by $g$, the environmental induced decoherence rate $A_{\rm heat}=\Gamma n_\text{th}$ acts in the opposite fashion. The pulse duration that leads to optimal entanglement is found to be on the order of $A_{\rm heat}^{-1}$ and the maximally achievable value of the logarithmic negativity can be estimated by
\begin{equation}
E_{\mathcal{N}}^{\rm max}\approx -{\rm ln}\left[1-2\left(\frac{g}{A_{\rm heat}}\right)^2\right].
\end{equation}
The result shows that a necessary condition for such a pulsed scheme to generate considerable entanglement is that the coupling strength overcomes the thermal reheating rate: this is validated by numerical simulations shown in Fig.~\ref{Fig2}(a). The robustness of such optoacoustic entanglement with respect to
environment temperature is presented in Fig.~\ref{Fig2}(b) showing that high values of optoacoustic entanglement, comparable
to other recent results~\cite{Yumei,Yafeng,Denggao}, can be achieved at a temperature
of tens of Kelvins. For example, using the setup of Ref~\cite{Laura}, for a waveguide with the length $L=0.5~$m and Brillouin gain $G_B=300~{\rm m}^{-1}{\rm W}^{-1}$,
it should be possible to create entangled photon-phonon pairs with $E_{\mathcal{N}}=0.3$ at a temperature of $30~$K by utilizing
a pulsed pump with duration of $11~$ns and peak power of $0.5~$W.

It should be noted that we only consider the case with a specific wavenumber $k$ in the
above discussion. However, the optical and acoustic waves provide groups
of photons and phonons in a continuous optomechanical
system~\cite{Kharel,Sipe,Hashem,Changlong,Junyin}. Therefore, the system has the capability of producing optoacoustic entanglement over a wide bandwidth of
Stokes photons and acoustic phonons.
This is illustrated in Fig.~\ref{Fig2}(c) which shows that the negativity $E_{\mathcal{N}}$ can achieve considerable values over a large interval of wavenumbers $k$, indicating high degree of entanglement over accessible groups of photons and phonons in Brillouin-active waveguides
at high environmental temperatures, which is a crucial aspects for a broad range of applications,
such as quantum computing~\cite{Andersen}, quantum communication~\cite{Saglamyurek}, and sensing~\cite{YiXia}.\\

\begin{figure}[t]
\centerline{
\includegraphics[width=9.0 cm, clip]{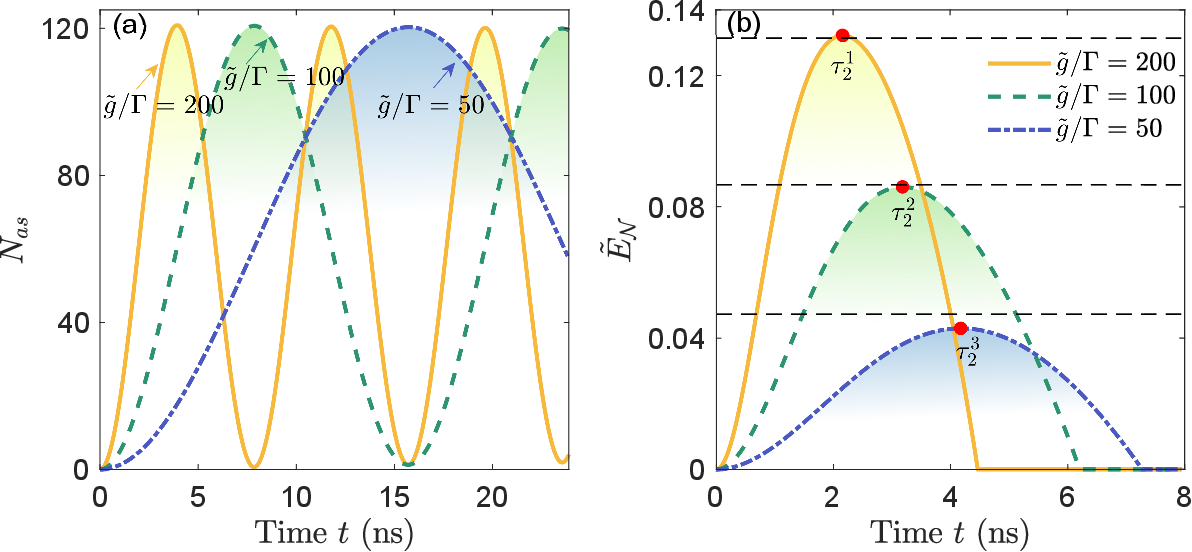}}
\caption{(a) Time evolution of the anti-Stokes photon occupation $n_\text{as}(t)$
in the strong coupling regime at $30~$K for various ratios $\tilde{g}/\Gamma$
during the optical readout process ($t\in \left[ \tau_1+\tau_d, \tau_1+\tau_d+\tau_2 \right]$).
(b) Corresponding bipartite entanglement between optical modes,
where red points denote the optimal time for optimal entanglement
and black dashed lines correspond to the maximum
value of $\tilde{E}_{\mathcal{N}}$ evaluated by the fully analytical expression.
The coupling strength in the entanglement process is fixed to $g/\Gamma=30$.
}\label{Fig3}
\end{figure}

\noindent \textbf{Optical readout of entanglement} ---
Although the detection of the optical Stokes field can be directly
performed by an optical heterodyne~\cite{Sarma,LiuQiu} or homodyne measurements~\cite{Vitali}, direct access to the quantum state of the acoustic field is
not easily experimentally achievable. This difficulty can be overcome by transducing the mechanical quantum state into the quantum state of an auxiliary optical mode, followed by standard homodyne detection on such mode. To this end, we propose to employ an additional backward-propagating pump, delayed by time $\tau_d$, as shown in Figs.~\ref{Fig1}(b) and (c). This maps the phonon state into anti-Stokes photons through Brillouin anti-Stokes scattering~\cite{Junyin}. In the undepleted pump regime, this process can be treated as a beam-splitter interaction. Such an interaction would lead to a full quantum state swap between the two modes; the additional presence of decoherence mechanisms simply degrades the fidelity of the swapping process thus requiring finite and quick swap pulses. In terms of quantum Langevin equations this can be cast as
\begin{eqnarray}\label{Motion equation in momentum spaceAS}
\frac{d \tilde{a}}{d t}&=& - \left( \frac{\gamma}{2} + i \Delta_{\tilde{a}} \right){\tilde{a}} - i \tilde{g} b + \sqrt{\gamma}\xi_{\tilde{a}}, \nonumber\\
\frac{d b}{d t}&=& - \left( \frac{\Gamma}{2} + i \Delta_b \right)b - i \tilde{g} {\tilde{a}} + \sqrt{\Gamma}\xi_b,
\end{eqnarray}
where ${\tilde{a}}$ is the bosonic operator referring at the anti-Stokes mode and we assumed that its loss rate is the same as the one of the Stokes mode $\gamma$. Notice that $\tilde g$ denotes the effective coupling strength between the anti-Stokes photons and acoustic phonons. The swapping dynamics can be simply understood from the analytical expression of the number of successfully transferred anti-Stokes quanta. In the
strong coupling regime, to a good approximaton, this is given by $e^{-(\gamma+\Gamma)t/2}
\left( 1-\cos{ 2\tilde{g}t} \right) n_{b}/2$,
where $t>0$ is the time counter during the readout pulse while $n_b$ corresponds to the
phonon number at the beginning of swapping time $\tau_1+\tau_d$. The result shows that an optimal swap is realized at  $t=\pi/(2\tilde{g})$
and affected by the exponential process stemming from optical and phonon decoherence.
The time evolution of $n_\text{as}(t)$ for various coupling strengths $\tilde{g}$ is illustrated in
Fig.~\ref{Fig3}(a).

Let us now move on onto the characterization of the observable entanglement between the Stokes and anti-Stokes modes. To this end we solve Eqs.~\eqref{Motion equation in momentum spaceAS} both analytically and numerically for the all optical logarithmic negativity
$\tilde{E}_{\mathcal{N}}$.
After some approximations, a simplified expression of the minimal symplectic eigenvalue
$\tilde{\lambda}_{-}$ can be casted in the following form
\begin{eqnarray}
\tilde{\lambda}_{-} \approx \frac{1}{2} \left[ 1 -\eta^2 t^2 + \frac{2}{3}\tilde{g}^2(\Gamma n_\text{th})  t^3 \right],
\end{eqnarray}
which resembles the solution obtained for the acoustic entanglement in Eq.~\eqref{Approximated_solution_of_lambda} up to third order in $t$ with an entanglement generation coefficient $\eta$ and the decoherence term identical as previously derived. The analytical expression for the entanglement readout rate is
\begin{eqnarray}
\eta^2 &=& \frac{\tilde{g}^2(\mathcal{C}_{ns}^2-4n_{b}n_{s})}{1+2n_{s}},
\end{eqnarray}
with notations $n_{s}$ and $\mathcal{C}_{ns}$ standing for the Stokes photon occupancy and
cross-correlation between Stokes photons and acoustic phonons at time $\tau_1+\tau_d$ (for more details see
Appendix). The solution is remarkably similar to the one in Eq.~\eqref{Approximated_solution_of_lambda} with the striking difference that the entanglement readout rate depends on the produced entanglement in the write-in part. This can be immediately observed in the expression $\mathcal{C}_{ns}^2-4n_{b}n_{s}$ indicating that cross-correlations are needed in order to swap entanglement. As a simple intuitive check, higher temperatures automatically reduce the swapping rate, as the product $n_{b}n_{s}$ increases, up to the point $\eta$ becomes negative and entanglement readout is no longer possible.

Numerical simulations are illustrated in Fig.~\ref{Fig3}(b) for various ratios of $\tilde{g}/\Gamma$,
where the red points correspond to the optimal pulse duration $\tau_2$ for various $\tilde{g}$, respectively.
At an environmental temperature around $30~$K, the thermal noise is considerable and needs a fast readout time as shown in Fig.~\ref{Fig3}(b).
Realistically, this can be obtained in a pulsed scheme where the entangling pump duration of $11~$ns and peak power of $0.5~$W
is followed by a counter-propagating pump with duration of $3~$ns and peak power of $5~$W
after $0.1~$ns delay. We assume a waveguide with the length $L=0.5~$m and Brillouin gain $G_B=300~{\rm m}^{-1}{\rm W}^{-1}$ as achieved in Ref.~\cite{Laura}. For an environmental temperature of $30~$K, the acoustic entanglement of peak value $\tilde{E}_{\mathcal{N}}=0.3$ will be transduced in an photon-photon entanglement of peak value  $\tilde{E}_{\mathcal{N}}=0.1$.

\noindent \textbf{Conclusions and outlook}--- We have shown that optoacoustic entanglement in continuous media, over groups
of photons and phonons modes can be achieved through the non-linear optical process of Brillouin scattering. As opposed to standard optomechanics, characterized by extremely high mechanical Q resonators and by operation under steady state conditions, the large mechanical loss characterizing such setups imply operation in the pulsed regime. This presents itself as an advantage, as the possibility of reaching strong phonon-photon coupling allows for the quick generation of entanglement, even in the presence of strong reheating rates. Our estimates show that, utilizing state-of-the-art
waveguides where continuous optomechanical cooling has been proved.~\cite{Laura}, the generated optoacoustic entanglement can survive in thermal environments far from the exclusive cryogenic requirements. Moreover, quantum ground state cooling of the acoustic mode in this case is not a prerequisite. The generated entanglement bimodal optoacoustic entanglement can be read out by optical detection means, namely by homodyne monitoring of both Stokes and anti-Stokes photons stemming from two different pumps which share the same propagating acoustic mode. The fact that the system operates over a large bandwidth of both optical and acoustic modes brings a new prospect of entanglement with continuum modes with great potential for applications in
quantum computation~\cite{Andersen}, quantum storage ~\cite{Saglamyurek}, quantum metrology~\cite{YiXia}, quantum teleportation~\cite{Barzanjeh}, entanglement-assisted quantum communication~\cite{Piveteau}, and the exploration of the boundary between classical and quantum worlds~\cite{Zurek}. \\

\noindent \textbf{ Acknowledgements} This work is supported by the Max-Planck-Society through the independent
Max Planck Research Groups Scheme and the Deutsche Forschungsgemeinschaft (DFG, German Research Foundation) -- Project-ID 429529648 -- TRR 306 QuCoLiMa (``Quantum Cooperativity of Light and Matter").

\newpage
\onecolumngrid

\appendix

\section{Dynamics of Brillouin Stokes scattering with undepleted pump}\label{S1}
In a typical Brillouin-active waveguide, the backward Brillouin Stokes scattering
provides an optomechanical interaction between a pump field, a scattered optical
field at redder Stokes frequency, and an acoustic field. The motion equations
of this Brillouin optoacoustic interaction can be given by
\begin{eqnarray}\label{Dynamics of Brillouin Stokes scattering}
  \frac{\partial a_\text{p}}{\partial t} + \upsilon_\text{opt}\frac{\partial a_\text{p}}{\partial z} &=&
  -\frac{\gamma}{2} a_\text{p} - i g_0 a_\text{s} b_\text{ac} + \sqrt{\gamma}\xi_\text{p},\nonumber\\
\frac{\partial a_\text{s}}{\partial t} - \upsilon_\text{opt}\frac{\partial a_\text{s}}{\partial z} &=&
  -\frac{\gamma}{2} a_\text{s} - i g_0 b_{\text{ac}}^{\dagger}a_\text{p} + \sqrt{\gamma}\xi_\text{s}, \nonumber\\
\frac{\partial b_\text{ac}}{\partial t} + \upsilon_\text{ac}\frac{\partial b_\text{ac}}{\partial z} &=&
  -\frac{\Gamma}{2}b_\text{ac} - i g_0 a_\text{s}^{\dagger} a_\text{p} + \sqrt{\Gamma}\xi_\text{ac},
\end{eqnarray}
where $a_\text{p}$, $a_\text{s}$, and $b_\text{ac}$ correspond to the envelope operators of pump
field, Stokes field, and acoustic field with carrier frequencies
$\omega_\text{p}, \omega_\text{s}, \Omega_\text{ac}$, respectively. $\upsilon_\text{opt}$ ($\gamma$) and $\upsilon_\text{ac}$
($\Gamma$) denote group velocities (dissipation rates) of optical and acoustic
fields. $g_0$ quantifies the coupling strength between these three fields at the single
quanta level. Without loss of generality, we take $g_0$ real and positive~\cite{Laer}
in the following discussion. $\xi_\text{p}$, $\xi_\text{s}$, and $\xi_\text{ac}$ represent Langevin noises
for these three fields, which obey the following statistical properties
\begin{eqnarray}
\langle \xi_\text{p}(t,z)\rangle &=& \langle \xi_\text{s}(t,z) = \langle \xi_\text{ac}(t,z)= 0, \nonumber\\
\langle \xi_\text{p}^{\dagger} (t_1,z_1)\xi_\text{p}(t_2,z_2)\rangle &=&
  \langle \xi_\text{s}^{\dagger}(t_1,z_1)\xi_\text{s}(t_2,z_2)\rangle = 0, \nonumber\\
\langle \xi_\text{ac}^{\dagger}(t_1,z_1)\xi_\text{ac}(t_2,z_2)\rangle &=& n_\text{th}\delta(t_1-t_2)\delta(z_1-z_2),
\end{eqnarray}
where $n_\text{th}=1/(e^{\hbar\Omega_\text{ac}/k_\text{B} T_\text{m}}-1)$ denotes the thermal phonon occupation
of the acoustic field at temperature $T_\text{m}$. Considering an undepleted pump field, the
three-field optoacoustic interaction can be treated as an effective interaction between
optical Stokes field and acoustic field with an effective coupling strength. Thus
Eq.~(\ref{Dynamics of Brillouin Stokes scattering}) can be reduced to
\begin{eqnarray}\label{Reduced dynamics of Brillouin Stokes scattering}
\frac{\partial a_\text{s}}{\partial t} - \upsilon_\text{opt}\frac{\partial a_\text{s}}{\partial z} &=&
  -\frac{\gamma}{2} a_\text{s} - i g b_\text{ac}^{\dagger} + \sqrt{\gamma}\xi_\text{s}, \nonumber\\
\frac{\partial b_\text{ac}}{\partial t} + \upsilon_\text{ac}\frac{\partial b_\text{ac}}{\partial z} &=&
  -\frac{\Gamma}{2}b_\text{ac} - i g a_\text{s}^{\dagger} + \sqrt{\Gamma}\xi_\text{ac},
\end{eqnarray}
where $g=g_0\sqrt{\langle a_\text{p}^{\dagger}a_\text{p}\rangle}$ is the pump-enhanced coupling strength
between Stokes photons and acoustic phonons. As $a_\text{s}$ and $b_\text{ac}$ represent envelope operators
with continuum modes and can be expressed in the discrete variable representation as follows~\cite{Kharel,Sipe,Hashem}
\begin{eqnarray}\label{Discrete picture}
a_\text{s}(t,z) &=& \frac{1}{\sqrt{2\pi}} \int a(t,k) e^{-ikz}d k,\nonumber\\
b_\text{ac}(t,z) &=& \frac{1}{\sqrt{2\pi}} \int b(t,k) e^{ikz}d k.
\end{eqnarray}
Substituting the above discrete form of envelope operators into Eq.~(\ref{Reduced dynamics of Brillouin Stokes scattering}),
the dynamics of the reduced optomechanical interaction can be re-expressed as
\begin{eqnarray}\label{Reduced dynamics of Stokes process in momentum space}
\frac{d a(t,k)}{dt} &=& -i k\upsilon_\text{opt} a - \frac{\gamma}{2}a - i g b^{\dagger} + \sqrt{\gamma}\xi_{a}(t,k), \nonumber\\
\frac{d b(t,k)}{dt} &=& -i k\upsilon_\text{ac} b - \frac{\Gamma}{2}b - i g a^{\dagger} + \sqrt{\Gamma}\xi_{b}(t,k),
\end{eqnarray}
where $a$ ($b$) represents the annihilation operator for the $k$-th Stokes photon (acoustic phonon)
mode with wavenumber $k$. $\xi_{a}(t,k)$ and $\xi_{b}(t,k)$ are inverse Fourier transform
of Langevin noises $\xi_\text{s}(t,z)$ and $\xi_\text{ac}(t,z)$, respectively. Thus the dynamics of the
reduced Brillouin Stokes process in momentum space can be described as follows
\begin{eqnarray}\label{Reduced dynamics of Stokes process in momentum space without subscript}
\frac{da}{dt}&=&-\left( \frac{\gamma}{2}+i\Delta_a \right)a - i g b^{\dagger} + \sqrt{\gamma}\xi_a, \nonumber\\
\frac{db}{dt}&=&-\left( \frac{\Gamma}{2}+i\Delta_b \right)b - i g a^{\dagger} + \sqrt{\Gamma}\xi_b,
\end{eqnarray}
where $\Delta_a = k\upsilon_\text{opt}$ and $\Delta_b = k\upsilon_\text{ac}$ denote wavenumber-induced frequency shifts
for Stokes and acoustic modes. $\xi_a$ and $\xi_b$ are zero-mean quantum Gaussian noises which obey
$\langle \xi_a(t_1)\xi_a^{\dagger}(t_2)\rangle=\delta(t_1-t_2)$ and
$\langle \xi_b(t_1)\xi_b^{\dagger}(t_2)\rangle=(1+n_\text{th})\delta(t_1-t_2)$.
The analytical solutions of Eq.~(\ref{Reduced dynamics of Stokes process in momentum space without subscript})
can be given by
\begin{eqnarray}\label{Analytical solution of Stokes process}
a(t) &=& \left( \mu_2 e^{\omega_{+} t} - \mu_3 e^{\omega_{-}t} \right) a_s(0)
  + \left( -e^{\omega_{+}t} + e^{\omega_{-}t} \right)\mu_1 b^{\dagger}(0)
  +\sqrt{\gamma} \int_0^t \left[ \mu_2 e^{\omega_{+}(t-\tau)} - \mu_3 e^{\omega_{-}(t-\tau)} \right] \xi_a(\tau) d\tau \nonumber\\
 &&+ \sqrt{\Gamma}\mu_1 \int_0^t \left[ -e^{\omega_{+}(t-\tau)} + e^{\omega_{-}(t-\tau)} \right] \xi_b^{\dagger}(\tau)d\tau, \nonumber\\
b(t) &=& \mu_1^{*} \left( e^{\omega_{+}^{*}t} - e^{\omega_{-}^{*}t} \right) a^{\dagger}(0)
  + \left( -\mu_3^{*} e^{\omega_{+}^{*}t} + \mu_2^{*} e^{\omega_{-}^{*}t} \right) b(0)
  + \sqrt{\gamma} \mu_1^{*} \int_0^t \left[ e^{\omega_{+}^{*}(t-\tau)} - e^{\omega_{-}^{*}(t-\tau)} \right]
  \xi_a^{\dagger}(\tau) d\tau \nonumber\\
 &&+ \sqrt{\Gamma} \int_0^t \left[ -\mu_3^{*} e^{\omega_{+}^{*}(t-\tau)} + \mu_2^{*} e^{\omega_{-}^{*}(t-\tau)}  \right] \xi_b(\tau)d\tau,
\end{eqnarray}
where
\begin{eqnarray}
\omega_{+} &=& -\frac{\gamma+\Gamma}{4} - i\frac{\Delta_a-\Delta_b}{2}
  - \frac{ \sqrt{ 16g^2 + \left[ (\Gamma-\gamma) + 2i(\Delta_a+\Delta_b) \right]^2 } }{4}, \nonumber\\
\omega_{-} &=& -\frac{\gamma+\Gamma}{4} - i\frac{\Delta_a-\Delta_b}{2}
  + \frac{ \sqrt{ 16g^2 + \left[ (\Gamma-\gamma) + 2i(\Delta_a+\Delta_b) \right]^2 } }{4}, \nonumber\\
\tau_{+} &=& -\frac{ 2(\Delta_a+\Delta_b)+i(\Gamma-\gamma) }{4g}
  + i\frac{ \sqrt{ 16g^2 + \left[ (\Gamma-\gamma) + 2i(\Delta_a+\Delta_b) \right]^2 } }{4g}, \nonumber\\
\tau_{-} &=& -\frac{ 2(\Delta_a+\Delta_b)+i(\Gamma-\gamma) }{4g}
  - i\frac{ \sqrt{ 16g^2 + \left[ (\Gamma-\gamma) + 2i(\Delta_a+\Delta_b) \right]^2 } }{4g},
\end{eqnarray}
and
\begin{eqnarray}
\mu_1 = \frac{1}{\tau_{+} - \tau_{-}}, \quad \mu_2 = \frac{\tau_{+}}{\tau{+}-\tau_{-}}, \quad
\mu_3 = \frac{\tau_{-}}{ \tau_{+} - \tau_{-} }
\end{eqnarray}

\section{Optoacoustic entanglement generated by Brillouin Stokes process  }\label{S2}
In the above discussion, we analyzed the dynamics of Brillouin Stokes process and obtained
the analytical solutions of the generated Stokes photons and acoustic phonons. In order to
quantify the inseparability between Stokes photons and acoustic phonons, we consider the
logarithmic negativity $E_{\mathcal{N}}$~\cite{Zyczkowski,Vidal,Adesso}.

We define the quadrature phase operators of Stokes photons and acoustic phonons as follows
\begin{eqnarray}\label{Definition of quadratures}
x_1 &=& \frac{a+a^{\dagger}}{\sqrt{2}}, \quad p_1 = \frac{a-a^{\dagger}}{i\sqrt{2}}, \nonumber\\
x_2 &=& \frac{b+b^{\dagger}}{\sqrt{2}}, \quad p_2 = \frac{b-b^{\dagger}}{i\sqrt{2}}.
\end{eqnarray}
By applying the following variable transformation
\begin{eqnarray}
\left[ \begin{array}{ccc}
\phi_1\\
\phi_2\\
\phi_3\\
\phi_4
\end{array}
\right] =
\left[ \begin{array}{ccc}
x_1\\
p_1\\
x_2\\
p_2
\end{array}
\right],
\end{eqnarray}
the covariance matrix $\mathcal{V}$ which evaluates the correlation between Stokes photons
and acoustic phonons can be expressed as
\begin{eqnarray}\label{Elements of covariance matrix}
\mathcal{V}_{ij} = \frac{ \langle \phi_i(t)\phi_j(t)\rangle + \langle \phi_j(t)\phi_i(t) \rangle }{2}
- \langle \phi_i(t) \rangle \langle \phi_j(t) \rangle.
\end{eqnarray}
This symmetric covariance matrix $\mathcal{V}$ can be also re-written in the $2\times 2$ block
form
\begin{eqnarray}
\mathcal{V} =
\left[  \begin{array}{ccc}
A & C\\
C^{T} & B
\end{array}
\right].
\end{eqnarray}
We define the logarithmic negativity $E_{\mathcal{N}}$~\cite{Vidal,Adesso} as follows
\begin{eqnarray}
E_{\mathcal{N}} = \max\left[ 0, -\ln2\lambda_{-} \right],
\end{eqnarray}
where $\lambda_{-}$ denotes the minimal symplectic eigenvalue of the matrix, which is
achieved by applying a partial transposition to the covariance matrix ~\cite{Pirandola}
and can be calculated as follows
\begin{eqnarray}\label{Minimal symplectic eigenvalue}
\lambda_{-}\equiv \frac{\sqrt{ \Sigma(\mathcal{V}) - \sqrt{ \Sigma(\mathcal{V})^2 - 4\det\mathcal{V} } } }
{\sqrt{2}},
\end{eqnarray}
with
\begin{eqnarray}
\Sigma(\mathcal{V}) = \det A + \det B -2\det C.
\end{eqnarray}
We assume that there is no optical Stokes field at the beginning, i.e., the vacuum state,
and the initial state of the acoustic field is the thermal state with phonon occupation
$n_{0}=n_\text{th}$. Substituting Eq.~(\ref{Analytical solution of Stokes process}) into
Eq.~(\ref{Elements of covariance matrix}) and combining with the initial states of
optical and acoustic modes and properties of Langevin noises $\xi_{a,b}$, the ten
independent elements of covariance matrix $\mathcal{V}$ can be expressed as
\begin{align}
\mathcal{V}_{11} &= \frac{ | \mu_2 e^{\omega_{+}t} - \mu_3 e^{\omega_{-}t} |^2 }{2}
  + \frac{ | -e^{\omega_{+}t} + e^{\omega_{-}t} |^2 }{2} |\mu_1|^2(2n_0+1) \nonumber\\
 &+ \frac{ \gamma|\mu_2|^2 + \Gamma(2n_\text{th}+1)|\mu_1|^2 }{2\alpha_1} ( e^{\alpha_1 t}-1 )
  - \frac{ \gamma\mu_2\mu_3^{*} + \Gamma(2n_\text{th}+1)|\mu_1|^2 }{2\alpha_2}(e^{\alpha_2 t}-1) \nonumber\\
 &- \frac{ \gamma\mu_3\mu_2^{*} + \Gamma(2n_\text{th}+1)|\mu_1|^2 }{2\alpha_3}(e^{\alpha_3 t}-1)
  + \frac{ \gamma|\mu_3|^2 + \Gamma(2n_\text{th}+1)|\mu_1|^2 }{2\alpha_4}(e^{\alpha_4 t}-1), \tag{17-1}\\
\mathcal{V}_{12} &= 0, \tag{17-2}\\
\mathcal{V}_{13} &= \frac{ ( e^{\omega_{+}t} - e^{\omega_{-}t} )
  ( \mu_2^{*}e^{\omega_{+}^{*}t} - \mu_3^{*}e^{\omega_{-}^{*}t} ) }{4}\mu_1
  + \frac{ ( \mu_2e^{\omega_{+}t} - \mu_3e^{\omega_{-}t} )( e^{\omega_{+}^{*}t} - e^{\omega_{-}^{*}t} ) }{4}\mu_1^{*} \nonumber\\
 &+ \frac{ ( \mu_1\mu_3^{*} + \mu_3\mu_1^{*} )e^{\alpha_1 t}
  - ( \mu_1\mu_2^{*} + \mu_3\mu_1^{*} )e^{\alpha_2 t}
  - ( \mu_3^{*}\mu_1 + \mu_2\mu_1^{*} )e^{\alpha_3 t}
  + ( \mu_1\mu_2^{*} + \mu_2\mu_1^{*} )e^{\alpha_4 t} }{4} (1+2n_0) \nonumber\\
 &+ \frac{ \gamma( \mu_1\mu_2^{*}+\mu_1^{*}\mu_2 )
  + \Gamma( 1+2n_\text{th} )( \mu_1\mu_3^{*}+\mu_3\mu_1^{*} ) }{4\alpha_1}(e^{\alpha_1 t}-1) \nonumber\\
 &- \frac{ \gamma( \mu_1\mu_3^{*} + \mu_1^{*}\mu_2 )
  + \Gamma( 1+2n_\text{th} )( \mu_1\mu_2^{*}+\mu_3\mu_1^{*} ) }{4\alpha_2}(e^{\alpha_2 t}-1) \nonumber\\
 &- \frac{ \gamma( \mu_1\mu_2^{*}+\mu_1^{*}\mu_3 )
  + \Gamma( 1+2n_\text{th} )( \mu_1\mu_3^{*} + \mu_2\mu_1^{*} ) }{4\alpha_3} (e^{\alpha_3 t}-1) \nonumber\\
 &+ \frac{ \gamma( \mu_1\mu_3^{*} + \mu_1^{*}\mu_3 )
  + \Gamma( 1+2n_\text{th} )( \mu_1\mu_2^{*}+\mu_2\mu_1^{*} ) }{4\alpha_4}( e^{\alpha_4 t} - 1 ), \tag{17-3}
\end{align}
\begin{align}
\mathcal{V}_{14} &= i\frac{ ( \mu_2^{*}\mu_1 - \mu_2\mu_1^{*} )e^{\alpha_1 t}
  + ( \mu_1^{*}\mu_2 - \mu_3^{*}\mu_1 )e^{\alpha_2 t}
  + ( \mu_1^{*}\mu_3 - \mu_2^{*}\mu_1 )e^{\alpha_3 t}
  + ( \mu_3^{*}\mu_1 - \mu_1^{*}\mu_3 )e^{\alpha_4 t} }{4} \nonumber\\
 &+ i\frac{ ( \mu_3\mu_1^{*} - \mu_1\mu_3^{*} )e^{\alpha_1 t}
  + ( \mu_1\mu_2^{*} - \mu_3\mu_1^{*} )e^{\alpha_2 t}
  + ( \mu_1\mu_3^{*} - \mu_2\mu_1^{*} )e^{\alpha_3 t}
  + ( \mu_2\mu_1^{*} - \mu_1\mu_2^{*} )e^{\alpha_4 t} }{4} (1+2n_0) \nonumber\\
 &+ i\frac{ \gamma( \mu_1\mu_2^{*} - \mu_1^{*}\mu_2 )+\Gamma( 1+2n_\text{th} )
  ( \mu_3\mu_1^{*}-\mu_1\mu_3^{*} ) }{4\alpha_1}(e^{\alpha_1 t}-1) \nonumber\\
 &+ i\frac{ \gamma( \mu_1^{*}\mu_2 - \mu_1\mu_3^{*} )
  +\Gamma( 1+2n_\text{th} )( \mu_1\mu_2^{*}-\mu_1^{*}\mu_3 ) }{4\alpha_2}(e^{\alpha_2 t}-1) \nonumber\\
 &+ i\frac{ \gamma( \mu_1^{*}\mu_3 - \mu_1\mu_2^{*} )
  +\Gamma(1+2n_\text{th})( \mu_1\mu_3^{*} - \mu_1^{*}\mu_2 ) }{4\alpha_3}(e^{\alpha_3 t}-1) \nonumber\\
 & + i\frac{ \gamma(\mu_1\mu_3^{*}-\mu_3\mu_1^{*})
  + \Gamma( 1+2n_\text{th} )( \mu_2\mu_1^{*} - \mu_1\mu_2^{*} ) }{4\alpha_4}(e^{\alpha_4 t}-1), \tag{17-4} \\
\mathcal{V}_{22} &= \mathcal{V}_{11}, \tag{17-5}\\
\mathcal{V}_{23} &= \mathcal{V}_{14}, \tag{17-6}\\
\mathcal{V}_{24} &= -\mathcal{V}_{13}, \tag{17-7}\\
\mathcal{V}_{33} &= \frac{ e^{\omega_{+}t} - e^{\omega_{-}t} }{2}|\mu_1|^2
  + \frac{ |\mu_3|^2e^{\alpha_1 t} - \mu_3\mu_2^{*}e^{\alpha_2 t}
  - \mu_2\mu_3^{*}e^{\alpha_3 t} + |\mu_2|^2 e^{\alpha_4 t} }{2}(1+2n_0) \nonumber\\
 &+ \frac{ \gamma|\mu_1|^2 + \Gamma(1+2n_\text{th})|\mu_3|^2 }{2\alpha_1}(e^{\alpha_1 t}-1)
  - \frac{ \gamma|\mu_1|^2 + \Gamma(1+2n_\text{th})\mu_3\mu_2^{*} }{2\alpha_2}(e^{\alpha_2 t}-1) \nonumber\\
 &- \frac{ \gamma|\mu_1|^2 + \Gamma(1+2n_\text{th})mu_2\mu_3^{*} }{2\alpha_3}(e^{\alpha_3 t}-1)
  + \frac{ \gamma|\mu_1|^2+\Gamma(1+2n_\text{th})|\mu_2|^2 }{2\alpha_4}(e^{\alpha_4 t}-1) \tag{17-8}\\
\mathcal{V}_{34} &= 0, \tag{17-9}\\
\mathcal{V}_{44} &= \mathcal{V}_{33}, \tag{17-10} \\ \label{Original elements of covariance matrix}
\end{align}
where
\begin{eqnarray}
\alpha_1 &=& \omega_{+} + \omega_{+}^{*}, \quad \alpha_2 = \omega_{+} + \omega_{-}^{*}, \nonumber\\
\alpha_3 &=& \omega_{-} + \omega_{+}^{*}, \quad \alpha_4 = \omega_{-} + \omega_{-}^{*}. \nonumber
\end{eqnarray}
Substituting above solutions into $\Sigma(\mathcal{V})$ and $\det\mathcal{V}$, we have
\begin{eqnarray}
\Sigma(\mathcal{V}) &=& \mathcal{V}_{11}^{2} + \mathcal{V}_{33}^{2} + 2\left(\mathcal{V}_{13}^2 + \mathcal{V}_{14}^2\right), \nonumber\\
\det\mathcal{V} &=& \left( \mathcal{V}_{13}^2 + \mathcal{V}_{14}^2 - \mathcal{V}_{11} \mathcal{V}_{33} \right)^2.
\end{eqnarray}
We consider the strong coupling regime, i.e., $g\gg \gamma, \Gamma$ and assume that the wavenumber-induced
frequency shifts $\Delta_{a,b}$ are within the linewidth of the acoustic field, i.e., $\Delta_{a,b}<\Gamma$.
For the backward Brillouin scattering in a typical waveguide, the optical loss is far smaller than the acoustic
loss, i.e., $\gamma\ll\Gamma$. Under these conditions, we have $\det\mathcal{V}/(\Sigma(\mathcal{V}))^2\ll1$
and thus can approximately calculate $\lambda_{-}$ as follows
\begin{eqnarray}
\lambda_{-} &=& \sqrt{ \frac{ \Sigma(\mathcal{V}) - \sqrt{ (\Sigma(\mathcal{V}))^2 - 4\det\mathcal{V} } }{2} } \nonumber\\
 &=& \sqrt{ \frac{ \Sigma(\mathcal{V}) - \Sigma(\mathcal{V})\sqrt{ 1 - \frac{4\det\mathcal{V}}{(\Sigma(\mathcal{V}))^2} } }{2} }\nonumber\\
 &\approx& \sqrt{ \frac{\det\mathcal{V}}{\Sigma(\mathcal{V})} } \nonumber\\
 &\approx& \frac{ \left| \mathcal{V}_{13}^2 + \mathcal{V}_{14}^2 - \mathcal{V}_{11} \mathcal{V}_{33} \right| }
   {\sqrt{ \mathcal{V}_{11}^{2} + \mathcal{V}_{33}^{2} + 2\left(\mathcal{V}_{13}^2 + \mathcal{V}_{14}^2\right) }} \nonumber\\
 &\approx& \frac{ \left| \mathcal{V}_{14}^2 - \mathcal{V}_{11} \mathcal{V}_{33} \right| }
  { \sqrt{ \mathcal{V}_{11}^{2} + \mathcal{V}_{33}^{2} + 2\mathcal{V}_{14}^2 } } \nonumber\\
 &\approx& \frac{ \left| \mathcal{V}_{14}^2 \mathcal{V}_{33} - \mathcal{V}_{11} \mathcal{V}_{33}^2 \right| }
  { \mathcal{V}_{33}^2 + \mathcal{V}_{14}^2  },
\end{eqnarray}
i.e.,
\begin{eqnarray}\label{Eigenvalue at 30K}
\lambda_{-} &\approx& \frac{ \left| \mathcal{V}_{14}^2 \mathcal{V}_{33} - \mathcal{V}_{11} \mathcal{V}_{33}^2 \right| }
  { \mathcal{V}_{33}^2 + \mathcal{V}_{14}^2  }.
\end{eqnarray}
The approximated elements of $\mathcal{V}_{33}$, $\mathcal{V}_{14}$, and $\mathcal{V}_{11}$ can be given by
\begin{eqnarray}\label{Elements at 30K}
\mathcal{V}_{33}&\approx& \eta_{11}e^{\alpha_1t} + \eta_{12}e^{\alpha_4t} + \eta_{13}e^{-\frac{\gamma+\Gamma}{2}t} + \eta_{14}, \nonumber\\
\mathcal{V}_{14}&\approx& \eta_{21}e^{\alpha_1t} + \eta_{22}e^{\alpha_4t} + \eta_{23}e^{-\frac{\gamma+\Gamma}{2}t} + \eta_{24}, \nonumber\\
\mathcal{V}_{11}&\approx& \eta_{31}e^{\alpha_1t} + \eta_{32}e^{\alpha_4t} + \eta_{33}e^{-\frac{\gamma+\Gamma}{2}t} + \eta_{34},
\end{eqnarray}
where the coefficients $\eta_{ij}$ can be calculated as
\begin{eqnarray}
\eta_{11} &=& \frac{g^2(1+n_\text{th})}{4\Delta} \left[ 1 + \frac{\Gamma-\gamma}{2g} -
  \frac{2\Gamma+\frac{\Gamma(\Gamma-\gamma)}{g}}{4\sqrt{\Delta}+(\gamma+\Gamma)} \right], \nonumber\\
\eta_{12} &=& \frac{g^2}{8\Delta} \left[ 1 + \frac{ (g-\frac{\Gamma-\gamma}{4})^2 + (\frac{\Gamma-\gamma}{4})^2 }{g^2}(1+2n_\text{th})
  + 2\frac{ (g-\frac{\Gamma-\gamma}{4})^2 + (\frac{\Gamma-\gamma}{4})^2 }{g^2\left( 4\sqrt{\Delta}-(\gamma+\Gamma) \right)}
  \Gamma(1+2n_\text{th}) \right], \nonumber\\
\eta_{13} &=& -\frac{ \Gamma+(\Gamma-\gamma)n_\text{th} }{2\Delta(\gamma+\Gamma)} g^2, \nonumber\\
\eta_{14} &=& (1+2n_\text{th})\Gamma g^2 \frac{ 16\Delta+(\Gamma+\gamma)(\Gamma-3\gamma) }{32\Delta^2(\Gamma+\gamma)}, \nonumber\\
\eta_{21} &=& \frac{g}{8\Delta} \left[ \frac{(\Gamma-\gamma)n_\text{th}}{2} + 2\sqrt{\Delta}(n_\text{th}+1)
  - \frac{\Gamma(1+2n_\text{th})}{2} \frac{4\sqrt{\Delta}+(\Gamma-\gamma)}{ 4\sqrt{\Delta}+(\Gamma+\gamma) }  \right], \nonumber\\
\eta_{22} &=& -\frac{g}{8\Delta} \left[ -\frac{(\Gamma-\gamma)n_\text{th}}{2} + 2\sqrt{\Delta}(n_\text{th}+1)
  + \frac{\Gamma(1+2n_\text{th})}{2} \frac{4\sqrt{\Delta}-(\Gamma-\gamma)}{ 4\sqrt{\Delta} - (\Gamma+\gamma) } \right], \nonumber\\
\eta_{23} &=& \frac{g}{8\Delta} \left[ -(\Gamma-\gamma)n_\text{th} + \frac{(\Gamma^2-\gamma^2)\Gamma(1+2n_\text{th})}{(\Gamma+\gamma)^2} \right], \nonumber\\
\eta_{24} &=& \frac{ \gamma\Gamma g (1+2n_\text{th}) }{4\Delta(\Gamma+\gamma)}, \nonumber\\
\eta_{31} &=& \frac{g^2}{8\Delta} \left[  2(n_\text{th}+1) - \frac{\Gamma-\gamma}{2g}
  - \frac{2\left( \gamma+\Gamma(2n_\text{th}+1) \right)}{4\sqrt{\Delta}+(\Gamma+\gamma)} \right], \nonumber\\
\eta_{32} &=& \frac{g^2}{8\Delta} \left[  2(n_\text{th}+1) + \frac{\Gamma-\gamma}{2g}
  + \frac{2\left( \gamma+\Gamma(2n_\text{th}+1) \right)}{4\sqrt{\Delta}+(\Gamma+\gamma)} \right], \nonumber\\
\eta_{33} &=& \frac{g^2(\Gamma-\gamma)(1+n_{th})}{2\Delta(\Gamma+\gamma)},  \nonumber\\
\eta_{34} &=& -\frac{ g^2 \left[ \Gamma(2n_\text{th}+1)-\gamma \right] \left[ 16\Delta+(\Gamma+\gamma)^2 \right] }{32\Delta^2(\Gamma+\gamma)},
\end{eqnarray}
with
\begin{eqnarray}
\alpha_1 &=& -\frac{\gamma+\Gamma}{2} - 2\sqrt{\frac{A+\Delta}{2}}, \nonumber\\
\alpha_4 &=& -\frac{\gamma+\Gamma}{2} + 2\sqrt{\frac{A+\Delta}{2}}, \nonumber\\
\Delta &=& g^2+\frac{(\Gamma-\gamma)^2}{16} - \frac{(\Delta_a+\Delta_b)^2}{4}, \nonumber\\
A &=& \sqrt{ \Delta^2 + \frac{(\Gamma-\gamma)^2(\Delta_a+\Delta_b)^2}{16} }.
\end{eqnarray}

Actually, if we consider the phase-matching condition ($\Delta_a=\Delta_b=0$),
the analytical expression of $\lambda_{-}$ after some simplifications can be reduced to
\begin{eqnarray}\label{Simplified expression of En with higher-orders of time}
\lambda_{-} &\approx& \frac{1}{2} \frac{X}{Y}, \nonumber \\
X &=& 1 + \frac{\Gamma(\Gamma n_\text{th} - g)}{2g} t + (g^2-\Gamma^2 n_\text{th}) t^2 + \frac{g(2g+3\Gamma)}{3} (\Gamma n_\text{th}) t^3
  + \frac{ g^2(g^2-4\Gamma^2 n_\text{th}) }{3} t^4 + \frac{4g^4(6g+5\Gamma)}{15(2g-\Gamma)} (\Gamma n_\text{th}) t^5 \nonumber\\
 && + \frac{2g^4(g^2-24\Gamma^2 n_\text{th})}{45} t^6 + \frac{g^6}{3}(\Gamma n_\text{th})t^7 + \frac{g^8}{315}t^8
  + \frac{4g^8}{45}(\Gamma n_\text{th}) t^9 + \frac{g^{10}}{50} (\Gamma n_\text{th})t^{11}, \nonumber\\
Y &=& 1 + \frac{3g(2g+\Gamma)}{2} t^2 + (2g+\Gamma)g^3 t^4 \left( \frac{4}{3} + \frac{11}{15}g^2t^2 +
  \frac{1}{5}g^4t^4+\frac{1}{26}g^6t^6 \right).
\end{eqnarray}

\begin{figure}[h]
\centerline{
\includegraphics[width=9 cm,clip]{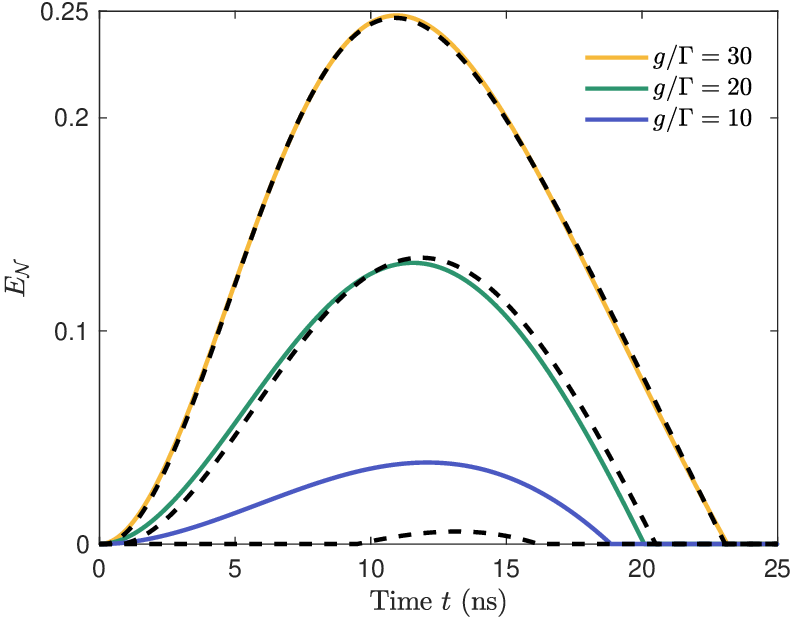}}
\caption{(Color online) Time evolution of logarithmic negativity $E_{\mathcal{N}}$ at temperature of $T_\text{m}=30~$K for
various ratios $g/\Gamma$, where solid curves denote the accurate simulation results of $E_{\mathcal{N}}$ and the
black dashed curves correspond to the approximated solutions evaluated by Eqs.~(\ref{Simplified expression of En with higher-orders of time}).
}\label{Sup_Fig1}
\end{figure}

We show the accurate simulations of $E_{\mathcal{N}}$ and the approximated results evaluated by
Eqs.~(\ref{Simplified expression of En with higher-orders of time}) in Fig.~\ref{Sup_Fig1}. We can
see that approximated results agree well with the accurate simulations when the effective
optoacoustic coupling strength is strong enough.

Equations~(\ref{Simplified expression of En with higher-orders of time}) are cumbersome and complicated.
Actually, we consider small evolution time to suppress the high thermal decoherence rate, i.e., $t\ll 1/\Gamma$. Thus the characteristic
behaviors of $\lambda_{-}$ can be predicted by a simplified expression as follows
\begin{eqnarray}\label{Simplified expression of En without higher-orders of time}
\lambda_{-} \approx \frac{1}{2} \times \frac{ 1 + g^2t^2 + \frac{2}{3}g^2(\Gamma n_\text{th})t^3 }{ 1 + 3g^2t^2 },
\end{eqnarray}
where higher-order terms of $t$ are ignored. If $gt\ll1$, Eq.~(\ref{Simplified expression of En without higher-orders of time})
can be further reduced to
\begin{eqnarray}
\lambda_{-} \approx \frac{1}{2} \left[ 1 - 2g^2t^2 + \frac{2}{3}g^2(\Gamma n_\text{th})t^3 \right].
\end{eqnarray}

\section{Entanglement at room temperature  }\label{S3}
\begin{figure}[h]
\centerline{
\includegraphics[width=14 cm,clip]{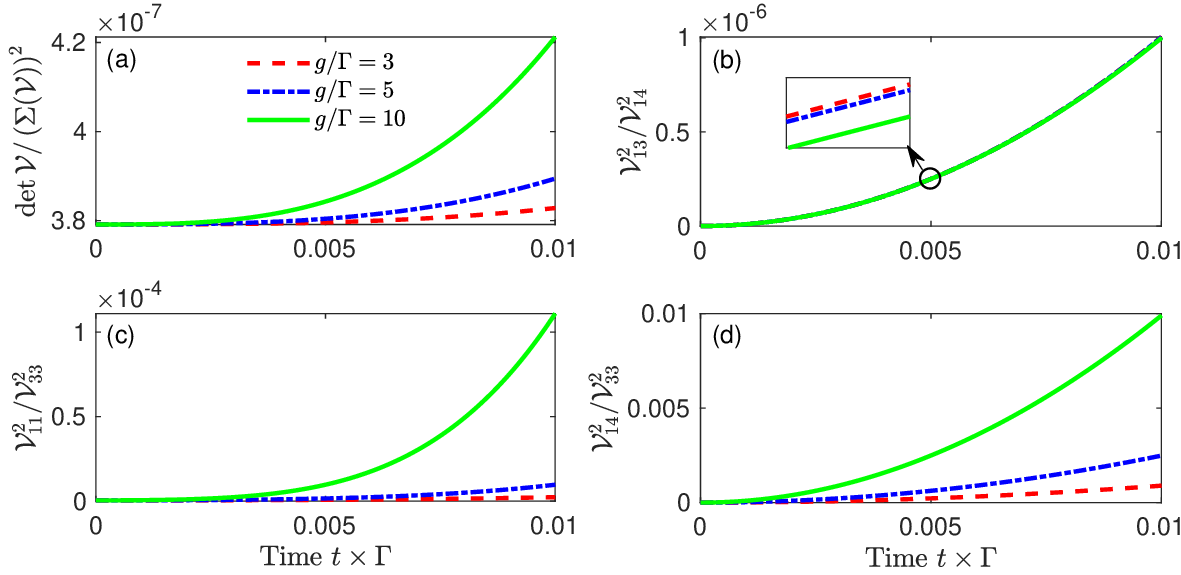}}
\caption{(Color online) Time evolution of (a) $\det\mathcal{V}/(\Sigma(\mathcal{V}))^2$
and (b) $\mathcal{V}_{13}^2/\mathcal{V}_{14}^2$ for $g/\Gamma=3, 5, 10$, where $t\ll\Gamma^{-1}$.
Other parameters are $\gamma/\Gamma=0.02$, $\Delta_a/\Gamma=0.4$,
$\Delta_b/\Gamma=4\times 10^{-5}$, $\Omega_\text{ac}=7.7$GHz and room temperature $T_\text{m}=300$K.
}\label{Sup_Fig2}
\end{figure}
In this section, we analyze the pulsed entanglement scheme for our system at room temperature ($300~$K).
We consider the strong coupling regime and thus have the relation
$\mathcal{V}_{13}^2\ll\mathcal{V}_{14}^2$ in our system when evolution time is
small ($t\ll\Gamma^{-1}$), as shown in Figs.~\ref{Sup_Fig2} (a) and (b). Thus $\lambda_{-}$ can be approximately
calculated as
\begin{eqnarray}
\lambda_{-} &=& \sqrt{ \frac{ \Sigma(\mathcal{V}) - \sqrt{ (\Sigma(\mathcal{V}))^2 - 4\det\mathcal{V} } }{2} } \nonumber\\
&=& \sqrt{ \frac{ \Sigma(\mathcal{V}) - \Sigma(\mathcal{V})\sqrt{ 1 - \frac{4\det\mathcal{V}}{(\Sigma(\mathcal{V}))^2} } }{2} }\nonumber\\
&\approx& \sqrt{ \frac{\det\mathcal{V}}{\Sigma(\mathcal{V})} } \nonumber\\
&\approx& \frac{ \left| \mathcal{V}_{13}^2 + \mathcal{V}_{14}^2 - \mathcal{V}_{11} \mathcal{V}_{33} \right| }
{\sqrt{ \mathcal{V}_{11}^{2} + \mathcal{V}_{33}^{2} + 2\left(\mathcal{V}_{13}^2 + \mathcal{V}_{14}^2\right) }} \nonumber\\
&\approx& \frac{ \left| \mathcal{V}_{14}^2 - \mathcal{V}_{11} \mathcal{V}_{33} \right| }
{ \sqrt{ \mathcal{V}_{11}^{2} + \mathcal{V}_{33}^{2} + 2\mathcal{V}_{14}^2 } }.
\end{eqnarray}

As $\mathcal{V}_{11}^2\ll\mathcal{V}_{33}^2$ and $\mathcal{V}_{14}^2\ll\mathcal{V}_{33}^2$ when $t\ll\Gamma^{-1}$,
as shown in Figs.~\ref{Sup_Fig2} (c) and (d), $\lambda_{-}$ can be further reduced to
\begin{eqnarray}\label{Minimum Eigenvalue with matrix elements}
\lambda_{-} &\approx& \frac{ \left| \mathcal{V}_{14}^2 - \mathcal{V}_{11}\mathcal{V}_{33} \right| }
{ \mathcal{V}_{33}\sqrt{ 1 + 2\frac{\mathcal{V}_{14}^2}{\mathcal{V}_{33}^{2}} } } \nonumber\\
&\approx& \frac{ \left| \mathcal{V}_{14}^2 - \mathcal{V}_{11}\mathcal{V}_{33} \right| }
{ \mathcal{V}_{33} + \frac{\mathcal{V}_{14}^2}{\mathcal{V}_{33}} } \nonumber\\
&\approx& \frac{ \left| \mathcal{V}_{33}\mathcal{V}_{14}^2 - \mathcal{V}_{11}\mathcal{V}_{33}^2 \right| }
{ \mathcal{V}_{33}^2 + \mathcal{V}_{14}^2 }.
\end{eqnarray}
This means that the minimal symplectic eigenvalue $\lambda{-}$ is mainly dependent on three independent
elements $\mathcal{V}_{11}$, $\mathcal{V}_{33}$, and $\mathcal{V}_{14}$. In fact, these three elements of
matrix $\mathcal{V}$ can be approximately expressed as
\begin{eqnarray}\label{Approximate expression of three independent elements}
\mathcal{V}_{11} &\approx& \frac{1}{4} \left[ 2 + (\Gamma-\gamma)t + \frac{16g^2(n_\text{th}+1)}{4}t^2
  + \frac{ 16g^2(n_\text{th}+1) }{ 12 }\Delta t^4 \right] e^{-\frac{\gamma+\Gamma}{2}t}
  + \frac{\gamma g^2}{2\Delta}t + \frac{ \Gamma g^2(2n_\text{th}+1) }{6}t^3, \nonumber\\
\mathcal{V}_{33} &\approx& \left( \frac{1}{2} - \frac{\Gamma-\gamma}{4}t + \frac{n_\text{th}+1}{2n_\text{th}+1}g^2t^2 \right)
  (2n_\text{th}+1) e^{-\frac{\gamma+\Gamma}{2}t} + \frac{ \Gamma(2n_\text{th}+1) }{2}t
  - \frac{ g^2\Gamma^2(2n_\text{th}+1) }{4\Delta}t^2, \nonumber\\
\mathcal{V}_{14} &\approx& \frac{g n_\text{th}}{4} \left[ -4(1+\frac{1}{n_\text{th}}) + (\Gamma-\gamma)t - \frac{8}{3}\Delta t^2 \right]
  t e^{ -\frac{\gamma+\Gamma}{2}t } + \frac{g\Gamma}{4}(1+2n_\text{th})\left( \frac{\Gamma^2}{16\Delta} - 1 \right) t^2,
\end{eqnarray}
where $\Delta = g^2 + \frac{(\Gamma-\gamma)^2}{16} - \frac{(\Delta_a+\Delta_b)^2}{4}$.
We substitute Eq.~(\ref{Approximate expression of three independent elements})
into Eq.~(\ref{Minimum Eigenvalue with matrix elements}) and use the first-order
approximation, i.e., $\exp{[-(\gamma+\Gamma)t/2]}\approx 1 - (\gamma+\Gamma)t/2$,
when the evolution time is short, then $\lambda_{-}$ can be approximately expressed as
\begin{eqnarray}\label{Reduced expression of lambda}
\lambda_{-} \approx \frac{1}{2} \left| \frac{ 1 + (g^2-\frac{\Gamma^2-\gamma^2}{4})t^2
  + \frac{g^2(3\Gamma^2+16g^2)}{24g^2}(\Gamma n_\text{th}) t^3 }{ 1 + 3g^2t^2 } \right|,
\end{eqnarray}
If $gt\ll1$, the above equation can be further simplified as follows
\begin{eqnarray}
\lambda_{-}\approx \frac{1}{2}\left[ 1 - 2g^2t^2 + \frac{2}{3}g^2(\Gamma n_\text{th})t^3 \right].
\end{eqnarray}

\begin{figure}[h]
\centerline{
\includegraphics[width=9 cm,clip]{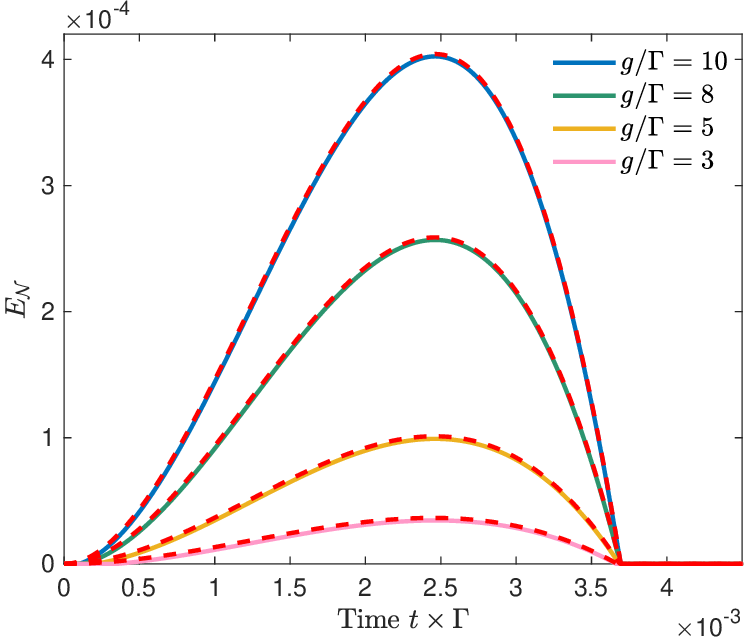}}
\caption{(Color online) Time evolution of $E_{\mathcal{N}}$ for different coupling strength
at room temperature $T_\text{m}=300~$K, where solid curves and red dashed curves correspond to the
accurate simulation results and approximated solutions evaluated by Eq.~(\ref{Reduced expression of lambda}),
respectively.
}\label{Sup_Fig3}
\end{figure}

Given the parameters used in Fig.~\ref{Sup_Fig2}, we show the time evolution of
$E_{\mathcal{N}}$ in the strong coupling regime in Fig.~\ref{Sup_Fig3}, where
the solid curves denote accurate simulation results based on Eq.~(\ref{Minimal symplectic eigenvalue})
and Eq.~(\ref{Original elements of covariance matrix}) and the dashed curves
correspond to the approximated solutions based on Eq.~(\ref{Reduced expression of lambda}).
Approximated results agree well with accurate analytical solutions. It shows that
similar to the case at $30~$K, there is a time window which allows optoacoustic
entanglement generation between Stokes photons and acoustic phonons in our system.

\section{State transfer via Brillouin anti-Stokes scattering  }\label{S4}
In the main text, we use the Brillouin anti-Stokes interaction to achieve the state transfer
between acoustic phonons and anti-Stokes photons. Now we present more details about this state
transfer in this section. For a typical Brillouin-active waveguide, the dynamics of the
backward Brillouin anti-Stokes scattering can be given by
\begin{eqnarray}\label{Intact dynamics of anti-Stokes process}
\frac{ \partial \tilde{a}_\text{p} }{\partial t} - \upsilon_\text{opt} \frac{ \partial \tilde{a}_\text{p} }{\partial z}
  &=& -\frac{\gamma}{2} \tilde{a}_\text{p} - i g_0^{*} \tilde{a}_\text{as} b_\text{ac}^{\dagger} + \sqrt{\gamma} \tilde{\xi}_\text{p}, \nonumber\\
\frac{ \partial \tilde{a}_\text{as} }{\partial t} + \upsilon_\text{opt} \frac{ \partial \tilde{a}_\text{as} }{\partial z}
  &=& -\frac{\gamma}{2} \tilde{a}_\text{as} - i g_0 \tilde{a}_\text{p} b_\text{ac} + \sqrt{\gamma}\tilde{\xi}_\text{as}, \nonumber\\
\frac{\partial b_\text{ac}}{\partial t} + \upsilon_\text{ac} \frac{\partial b_\text{ac}}{\partial z}
  &=& -\frac{\Gamma}{2} b_\text{ac} - i g_0 \tilde{a}_\text{p}^{\dagger} \tilde{a}_\text{as} + \sqrt{\Gamma} \xi_\text{ac},
\end{eqnarray}
where $\tilde{a}_\text{p}$ is the envelope operator of the second pump laser and used to stimulate the
anti-Stokes Brillouin interaction. $b_\text{ac}$ denotes the envelope operator of the
acoustic field generated during the entanglement generation process.
$\tilde{a}_\text{as}$ represents the envelope operator of the anti-Stokes field. $\tilde{\xi}_\text{p}$
and $\tilde{\xi}_{as}$ correspond to the Langevin noises.
Here we choose the propagation direction of the acoustic field as the positive direction of axis $z$.
Considering an undepleted pump, this three-field interaction can be reduced to an effective
optoacoustic interaction between anti-Stokes photons and acoustic phonons and thus
Eq.~(\ref{Intact dynamics of anti-Stokes process}) can be re-written as
\begin{eqnarray}
\frac{ \partial \tilde{a}_\text{as} }{\partial t} + \upsilon_\text{opt} \frac{ \partial \tilde{a}_\text{as} }{\partial z}
  &=& -\frac{\gamma}{2} \tilde{a}_\text{as} - i \tilde{g} b_\text{ac} + \sqrt{\gamma}\tilde{\xi}_\text{as}, \nonumber\\
\frac{\partial b_\text{ac}}{\partial t} + \upsilon_\text{ac} \frac{\partial b_\text{ac}}{\partial z}
  &=& -\frac{\Gamma}{2} b_\text{ac} -  i \tilde{g} \tilde{a}_\text{as} + \sqrt{\Gamma} \xi_\text{ac},
\end{eqnarray}
where $\tilde{g}=g_0\sqrt{\langle \tilde{a}_\text{p}^{\dagger}\tilde{a}_\text{p}\rangle}$ is the
pump-enhanced coupling strength which is tuned by the pump power. Actually, similar to
the acoustic field $b_\text{ac}$ in Eq.~(\ref{Discrete picture}), the anti-Stokes field $\tilde{a}_\text{as}$
denotes the envelope operator with continuum modes and can be expressed in the discrete variable
representation, i.e., $\tilde{a}_\text{as} = 1/\sqrt{2\pi}\int \tilde{a}(t,k) e^{ikz}dk$, where $\tilde{a}(t,k)$
corresponds to the annihilation operator for the $k$-th anti-Stokes photon mode with wavenumber $k$.
Moving into the momentum space, the dynamics of this linearized anti-Stokes Brillouin interaction
can be given by
\begin{eqnarray}\label{Simplified dynamics of anti-Stokes process in momentum space}
\frac{d b}{dt} &=& -(\frac{\Gamma}{2} + i\Delta_b)b - i \tilde{g} \tilde{a} + \sqrt{\Gamma}\xi_b, \nonumber\\
\frac{d \tilde{a}}{dt} &=& -(\frac{\gamma}{2} + i\Delta_{\tilde{a}})\tilde{a} - i \tilde{g} b + \sqrt{\gamma}\xi_{\tilde{a}},
\end{eqnarray}
where $\Delta_{\tilde{a}}=k\upsilon_\text{opt}$ is the wavenumber-induced frequency shift for anti-Stokes photons and $\xi_{\tilde{a}}$
is the Fourier transform of Langevin noise $\tilde{\xi}_\text{as}$ and a zero-mean quantum Gaussian noise with
$\langle \xi_{\tilde{a}}(t_1)\xi_{\tilde{a}}^{\dagger}(t_2)\rangle=\delta(t_1-t_2)$. Eqs.~(\ref{Simplified dynamics of anti-Stokes process in momentum space})
are ordinary differential equations with constant coefficients, which can be analytically solved.
Thus the analytical expression of the transferred anti-Stokes mode during time period $t\in\left[\Delta \tau, \Delta \tau + \tau_2 \right]$ can be given by
\begin{eqnarray}\label{Analytical expression of anti-Stokes mode}
\tilde{a}(t) &=& \left( \tilde{\mu}_2e^{\tilde{\omega}_{+}(t-\Delta\tau)} - \tilde{\mu}_3e^{\tilde{\omega}_{-}(t-\Delta\tau)} \right)\tilde{a}(\Delta\tau)
  + \tilde{\mu}_1 \left( e^{\tilde{\omega}_{+} (t-\Delta\tau)} - e^{\tilde{\omega}_{-} (t-\Delta\tau)} \right)b(\Delta\tau) \nonumber \\
 &&+ \sqrt{\gamma} \int_{\Delta\tau}^{t} \left[ \tilde{\mu}_2 e^{\tilde{\omega}_{+}(t-\tau)} - \tilde{\mu}_3 e^{\tilde{\omega}_{-}(t-\tau)} \right]\xi_{\tilde{a}}(\tau) d\tau
  + \sqrt{\Gamma}\tilde{\mu}_1 \int_{\Delta\tau}^{t} \left[  e^{\tilde{\omega}_{+}(t-\tau)} - e^{\tilde{\omega}_{-}(t-\tau)} \right] \xi_b(\tau) d\tau,
\end{eqnarray}
with
\begin{eqnarray}
\tilde{\omega}_{+} &=& -\frac{\gamma+\Gamma}{4} - i\frac{\Delta_b+\Delta_{\tilde{a}}}{2}
  - i\frac{ \sqrt{ 16\tilde{g}^2 - \left( (\Gamma-\gamma) - 2i(\Delta_{\tilde{a}}-\Delta_b) \right)^2 } }{4},  \nonumber\\
\tilde{\omega}_{-} &=& -\frac{\gamma+\Gamma}{4} - i\frac{\Delta_b+\Delta_{\tilde{a}}}{2}
  + i\frac{ \sqrt{ 16\tilde{g}^2 - \left( (\Gamma-\gamma) - 2i(\Delta_{\tilde{a}}-\Delta_b) \right)^2 } }{4},  \nonumber\\
\tilde{\tau}_{+} &=& \frac{ 2(\Delta_{\tilde{a}} - \Delta_b) + i(\Gamma-\gamma) }{4\tilde{g}}
  + \frac{ \sqrt{ 16\tilde{g}^2 - \left( (\Gamma-\gamma) - 2i(\Delta_{\tilde{a}}-\Delta_b) \right)^2 } }{4\tilde{g}},  \nonumber\\
\tilde{\tau}_{-} &=& \frac{ 2(\Delta_{\tilde{a}} - \Delta_b) + i(\Gamma-\gamma) }{4\tilde{g}}
  - \frac{ \sqrt{ 16\tilde{g}^2 - \left( (\Gamma-\gamma) - 2i(\Delta_{\tilde{a}}-\Delta_b) \right)^2 } }{4\tilde{g}},
\end{eqnarray}
and
\begin{eqnarray}
\tilde{\mu}_1 = \frac{1}{\tilde{\tau}_{+}-\tilde{\tau}_{-}}, \quad \tilde{\mu}_2 = \frac{\tilde{\tau}_{+}}{\tilde{\tau}_{+}-\tilde{\tau}{-}}, \quad \tilde{\mu}_3 = \frac{\tilde{\tau}_{-}}{ \tilde{\tau}_{+}-\tilde{\tau}_{-} },
\end{eqnarray}
where $\Delta\tau = \tau_1+\tau_d$.
Here we assume that there is no anti-Stokes photon present at initial time, i.e., $N_{as}(\Delta\tau)=0$.
Equation~(\ref{Analytical expression of anti-Stokes mode}) indicates that the generated anti-Stokes mode
follows the dynamics of the acoustic mode, i.e., the state can be transferred from the existed acoustic
phonons to the generated anti-Stokes photons. In order to investigate the influence of the thermal noise
on such phonon-photon state transfer, we calculate the generated mean photon number for the anti-Stokes
mode, which can be given by
\begin{eqnarray}\label{Completed expression for tranferred photons}
N_{as} (t) = \langle \tilde{a}^{\dagger}(t) \tilde{a}(t) \rangle
  = |\tilde{\mu}_1|^2 \left| e^{\tilde{\omega}_{+} (t-\Delta\tau)} - e^{\tilde{\omega}_{-} (t-\Delta\tau)} \right|^2 N_b(\Delta\tau)
  + \Gamma |\tilde{\mu}_1|^2 n_\text{th} \int_{\Delta\tau}^{t} \left| e^{\tilde{\omega}_{+} (t-\tau)} - e^{\tilde{\omega}_{-} (t-\tau)} \right|^2 d\tau.
\end{eqnarray}
The transferred anti-Stokes photon occupancy in the strong coupling regime can be simplified as follows
\begin{eqnarray}
N_{as}(t) \approx \frac{1}{2} e^{-\frac{\gamma+\Gamma}{2}(t-\Delta\tau)} \left[ 1 - \cos\left(2\tilde{g}(t-\Delta\tau) \right) \right]n_b,
\end{eqnarray}
where $n_b=N_b(\Delta\tau)$ corresponds to the the initial phonon occupation for the readout process.
Here we assume that the evolution time of the anti-Stokes process is short. The Rabi-oscillation property of $N_{as}$ implies that
the quantum states can be transferred between acoustic phonons and anti-Stokes photons via the anti-Stokes process.

For the entanglement readout process in the main text, we assume that the origin of time is taken
at $\tau_1+\tau_d$, i.e., $t\in [0,\tau_1+\tau_d]$, the transferred anti-Stokes photon occupation
can be re-written as
 \begin{eqnarray}
N_{as}(t) \approx \frac{1}{2} e^{-\frac{\gamma+\Gamma}{2}t} \left[ 1 - \cos\left(2\tilde{g}t \right) \right]n_b,
\end{eqnarray}
where $n_b$ denotes the initial Stokes photon occupation, i.e., $n_b=N_b(\tau_1+\tau_d)$.

\section{Entanglement measurement  }\label{S5}
As the quantum state of the acoustic phonons can be transferred to the anti-Stokes photons via the Brillouin
anti-Stokes process, the entanglement between the Stokes photons and acoustic phonons can be demonstrated
by quantifying the inseparability between the Stokes and anti-Stokes photons. In this subsection, we discuss
how to calculate the logarithmic negativity $\tilde{E}_{\mathcal{N}}$ between the optical Stokes field and
transferred optical anti-Stokes field during the entanglement measurement period, i.e., $t\in[\tau_1+\tau_d, \tau_1+\tau_d+\tau_2]$.

According to Eq.~(\ref{Analytical solution of Stokes process}), the Stoke and acoustic modes at time
$t=\tau_1$ can be given by
\begin{eqnarray}
a(\tau_1) &=& \left( \mu_2e^{\omega_{+}\tau_1} - \mu_3 e^{\omega_{-}\tau_1} \right) a(0)
  + \left( -e^{\omega_{+}\tau_1} + e^{\omega_{-}\tau_1} \right) \mu_1 b^{\dagger}(0)
  + \sqrt{\gamma} \int_0^{\tau_1} \left[ \mu_2 e^{\omega_{+}(\tau_1-\tau)} - \mu_3 e^{\omega_{-}(\tau_1-\tau)} \right]\xi_{a}(\tau)d\tau \nonumber\\
 && + \sqrt{\Gamma}\mu_1\int_0^{\tau_1} \left[ -e^{\omega_{+}(\tau_1-\tau)} + e^{\omega_{-}(\tau_1-\tau)} \right]\xi_b^{\dagger}(\tau)d\tau, \nonumber\\
b(\tau_1) &=& \mu_1^{*}\left( e^{\omega_{+}^{*}\tau_1} - e^{\omega_{-}^{*}\tau_1} \right)a^{\dagger}(0)
  + \left( -\mu_3^{*} e^{\omega_{+}^{*}\tau_1} + \mu_2^{*} e^{\omega_{-}^{*}\tau_1}  \right)b(0)
  + \sqrt{\gamma}\int_0^{\tau_1} \left[ e^{\omega_{+}^{*}(\tau_1-\tau)} - e^{\omega_{-}^{*}(\tau_1-\tau)} \right]\xi_a^{\dagger}(\tau)d\tau \nonumber\\
 && + \sqrt{\Gamma}\int_0^{\tau_1}\left[ -\mu_3^{*} e^{\omega_{+}^{*}(\tau_1-\tau)} + \mu_2^{*}e^{\omega_{-}^{*}(\tau_1-\tau)} \right]\xi_b(\tau)d\tau.
\end{eqnarray}
In addition, we assume that the output Stokes field produced during the entanglement generation process is fed into a single mode fiber and the corresponding
dynamics can be given by
\begin{eqnarray}
\frac{da}{dt} = -\left( \frac{\tilde{\gamma}}{2} + i\Delta_a \right)a_s + \sqrt{\tilde{\gamma}} \tilde{\xi}_{a}(t),
\end{eqnarray}
where $\tilde{\gamma}$ is the optical loss rate of the Stokes field in the single mode fiber and $\tilde{\xi}_a$ is the Langevin noise.
The analytical solution of the above equation can be expressed as
\begin{eqnarray}
a(t) = a(\tau_1) e^{-(\tilde{\gamma}/2+i\Delta_a)(t-\tau_1)}
  + \sqrt{\tilde{\gamma}}\int_{\tau_1}^{t} e^{-(\tilde{\gamma}/2+i\Delta_a)(t-\tau))} \tilde{\xi}_a(\tau) d\tau.
\end{eqnarray}
Furthermore, the acoustic field in the sample during the delay time period, i.e., $t\in[\tau_1,\tau_1+\tau_d]$, is driven by
the thermal noise and its dynamics can be given by
\begin{eqnarray}
\frac{db}{dt} = -\left( \frac{\Gamma}{2} + i\Delta_b \right) b + \sqrt{\Gamma}\xi_b.
\end{eqnarray}
The analytical solution of $b$ can be expressed as
\begin{eqnarray}
b(t) = b(\tau_1)e^{-(\Gamma/2+i\Delta_b)(t-\tau_1)}
  + \sqrt{\Gamma}\int_{\tau_1}^{t} e^{-(\Gamma/2+i\Delta_b)(t-\tau)}\xi_b(\tau)d\tau.
\end{eqnarray}
Therefore, the initial states for the entanglement measurement process, i.e., at time $t=\tau_1+\tau_d$, can be expressed as
\begin{eqnarray}\label{Initial_states_for_entanglement_measure}
a(\tau_1+\tau_d) &=& \left( \mu_2e^{\omega_{+}\tau_1} - \mu_3e^{\omega_{-}\tau_1} \right) e^{-(\tilde{\gamma}/2+i\Delta_a)\tau_d} a_s(0)
  + \left( -e^{\omega_{+}\tau_1} + e^{\omega_{-}\tau_1} \right)\mu_1 e^{-(\tilde{\gamma}/2+i\Delta_a)\tau_d} b^{\dagger}(0) \nonumber\\
 && + \sqrt{\gamma} e^{-(\tilde{\gamma}/2+i\Delta_a)\tau_d} \int_0^{\tau_1} \left( \mu_2e^{\omega_{+}(\tau_1-\tau)}
  - \mu_3e^{\omega_{-}(\tau_1-\tau)} \right) \xi_a(\tau) d \tau \nonumber\\
 && + \sqrt{\Gamma}\mu_1e^{-(\tilde{\gamma}/2+i\Delta_a)\tau_d} \int_0^{\tau_1} \left( -e^{\omega_{+}(\tau_1-\tau)}
  + e^{\omega_{-}(\tau_1-\tau)} \right) \xi_b^{\dagger}(\tau) d \tau \nonumber\\
 && + \sqrt{\tilde{\gamma}}e^{-(\tilde{\gamma}/2+i\Delta_a)(\tau_1+\tau_d)} \int_{\tau_1}^{\tau_1+\tau_d} e^{(\tilde{\gamma}/2+i\Delta_a)\tau} \tilde{\xi}_a(\tau) d\tau, \nonumber \\
b(\tau_1+\tau_d) &=& \mu_1^{*}\left( e^{\omega_{+}^{*}\tau_1} - e^{\omega_{-}^{*}\tau_1} \right) e^{-(\Gamma/2+i\Delta_b)\tau_d} a^{\dagger}(0)
  + \left( -\mu_3^{*}e^{\omega_{+}^{*}\tau_1} + \mu_2^{*}e^{\omega_{-}^{*}\tau_1} \right)e^{-(\Gamma/2+i\Delta_b)\tau_d} b(0) \nonumber\\
 && + \sqrt{\gamma}\mu_1^{*} e^{-(\Gamma/2+i\Delta_b)\tau_d} \int_0^{\tau_1} \left( e^{\omega_{+}^{*}(\tau_1-\tau)} - e^{\omega_{-}^{*}(\tau_1-\tau)} \right)\xi_a^{\dagger}(\tau)d\tau \nonumber\\
 && + \sqrt{\Gamma} e^{-(\Gamma/2+i\Delta_b)\tau_d} \int_0^{\tau_1} \left( -\mu_3^{*}e^{\omega_{+}^{*}(\tau_1-\tau)}
  + \mu_2^{*}e^{\omega_{-}^{*}(\tau_1-\tau)} \right) \xi_b(\tau) d\tau \nonumber\\
 && + \sqrt{\Gamma}e^{-(\Gamma/2+i\Delta_b)(\tau_1+\tau_d)} \int_{\tau_1}^{\tau_1+\tau_d} e^{(\Gamma/2+i\Delta_b)\tau}\xi_b(\tau)d\tau.
\end{eqnarray}
The state of the anti-Stokes mode $\tilde{a}$ at time $t=\tau_1+\tau_d$ is the vacuum state. Thus the optical Stokes mode $a(t)$ and anti-Stokes
mode $\tilde{a}(t)$ during time $t\in[\Delta\tau, \Delta\tau+\tau_2]$ can be given by
\begin{eqnarray}\label{Solution of Stokes and anti-Stokes photons during measurement}
a(t) &=& a(\Delta\tau) e^{-(\tilde{\gamma}/2+i\Delta_a)(t-\Delta\tau)}
  + \sqrt{\tilde{\gamma}} \int_{\Delta\tau}^{t} e^{-(\tilde{\gamma}/2+i\Delta_a)(t-\tau)} \tilde{\xi}_a(\tau) d\tau, \nonumber\\
\tilde{a}(t) &=& \left( \tilde{\mu}_2 e^{\tilde{\omega}_{+}(t-\Delta\tau)} - \tilde{\mu}_3 e^{\tilde{\omega}_{-}(t-\Delta\tau)} \right)
  a(\Delta\tau) + \tilde{\mu}_1 \left( e^{\tilde{\omega}_{+}(t-\Delta\tau)} - e^{\tilde{\omega}_{-}(t-\Delta\tau)} \right) b(\Delta\tau) \nonumber\\
 && + \sqrt{\gamma}\int_{\Delta\tau}^{t} \left( \tilde{\mu}_2 e^{\tilde{\omega}_{+}(t-\tau)}
  - \tilde{\mu}_3 e^{\tilde{\omega}_{-}(t-\tau)} \right) \xi_{\tilde{a}}(\tau) d\tau
  + \sqrt{\Gamma}\tilde{\mu}_1 \int_{\Delta\tau}^{t} \left( e^{\tilde{\omega}_{+}(t-\tau)} - e^{\tilde{\omega}_{-}(t-\tau)} \right) \xi_b(\tau) d\tau.
\end{eqnarray}

Now we utilize the logarithmic negativity $\tilde{E}_{\mathcal{N}}$ to quantify the inseparability between the Stokes mode $a$ and
anti-Stokes mode $\tilde{a}$ during the entanglement measurement process. We define the quadrature operators as follows
\begin{eqnarray}
x_1 &=& \frac{a + a^{\dagger}}{\sqrt{2}}, \quad p_1 = \frac{a - a^{\dagger}}{i\sqrt{2}}, \nonumber\\
x_3 &=& \frac{\tilde{a} + \tilde{a}^{\dagger}}{\sqrt{2}}, \quad p_3 = \frac{\tilde{a}-\tilde{a}^{\dagger}}{i\sqrt{2}}.
\end{eqnarray}
The corresponding covariance matrix $\tilde{\mathcal{V}}$ can be given by
\begin{eqnarray}\label{Define of En2}
\tilde{\mathcal{V}}_{i,j} = \frac{ \langle \phi_i(t)\phi_j(t) + \phi_i(t)\phi_j(t) \rangle }{2}
- \langle \phi_i(t) \rangle \langle \phi_j(t) \rangle,
\end{eqnarray}
where $\phi^{T}(t) = \left( x_1(t), p_1(t), x_3(t), p_3(t) \right)$. Substituting Eqs.~(\ref{Solution of Stokes and anti-Stokes photons during measurement}) into
Eq.~(\ref{Define of En2}), the symmetrical covariance matrix $\tilde{E}_{\mathcal{N}}$ can be given by
\begin{eqnarray}
\tilde{\mathcal{V}} =
\left[  \begin{array}{cccc}
\tilde{\mathcal{V}}_{11} & 0 & \tilde{\mathcal{V}}_{13} & \tilde{\mathcal{V}}_{14} \\
0 & \tilde{\mathcal{V}}_{11} & \tilde{\mathcal{V}}_{14} & -\tilde{\mathcal{V}}_{13} \\
\tilde{\mathcal{V}}_{13} & \tilde{\mathcal{V}}_{14} & \tilde{\mathcal{V}}_{33} & 0 \\
\tilde{\mathcal{V}}_{14} & -\tilde{\mathcal{V}}_{13} & 0 & \tilde{\mathcal{V}}_{33}
\end{array}
\right],
\end{eqnarray}
where
\begin{eqnarray}
\tilde{\mathcal{V}}_{11} &=& \frac{ 1 - e^{-\tilde{\gamma}(t-\Delta\tau)} }{2}
  + \frac{ e^{-\tilde{\gamma}(t-\Delta\tau)} }{2} \left( \langle a(\Delta\tau)^{\dagger} a(\Delta\tau) \rangle
  + \langle a(\Delta\tau) a^{\dagger}(\Delta\tau)  \rangle \right), \nonumber\\
\tilde{\mathcal{V}}_{13} &=& e^{-(\tilde{\gamma}/2+i\Delta_a)(t-\Delta\tau)} \left( e^{\tilde{\omega}_{+}(t-\Delta\tau)}
  - e^{\tilde{\omega}_{-}(t-\Delta\tau)} \right) \tilde{\mu}_1 \frac{ \langle a(\Delta\tau)b(\Delta\tau) \rangle
  + \langle b(\Delta\tau)a(\Delta\tau) \rangle }{4} \nonumber\\
 && + e^{-(\tilde{\gamma}/2-i\Delta_a)(t-\Delta\tau)} \left( e^{\tilde{\omega}_{+}^{*}(t-\Delta\tau)}
  - e^{\tilde{\omega}_{-}^{*}(t-\Delta\tau)} \right) \tilde{\mu}_1^{*}
  \frac{ \langle a^{\dagger}(\Delta\tau)b^{\dagger}(\Delta\tau) \rangle
  + \langle b^{\dagger}(\Delta\tau)a^{\dagger}(\Delta\tau) \rangle }{4}, \nonumber\\
\tilde{\mathcal{V}}_{14} &=& -i e^{-(\tilde{\gamma}/2+i\Delta_a)(t-\Delta\tau)} \left( e^{\tilde{\omega}_{+}(t-\Delta\tau)}
  - e^{\tilde{\omega}_{-}(t-\Delta\tau)} \right) \tilde{\mu}_1 \frac{ \langle a(\Delta\tau)b(\Delta\tau) \rangle
  + \langle b(\Delta\tau)a(\Delta\tau) \rangle }{4} \nonumber\\
 && + i e^{-(\tilde{\gamma}/2-i\Delta_a)(t-\Delta\tau)} \left( e^{\tilde{\omega}_{+}^{*}(t-\Delta\tau)}
  - e^{\tilde{\omega}_{-}^{*}(t-\Delta\tau)} \right) \tilde{\mu}_1^{*}
  \frac{ \langle a^{\dagger}(\Delta\tau)b^{\dagger}(\Delta\tau) \rangle
  + \langle b^{\dagger}(\Delta\tau)a^{\dagger}(\Delta\tau) \rangle }{4},  \nonumber\\
\tilde{\mathcal{V}}_{33} &=& \frac{ \left| \tilde{\mu}_2 e^{\tilde{\omega}_{+}(t-\Delta\tau)}
  - \tilde{\mu}_3 e^{\tilde{\omega}_{-}(t-\Delta\tau)} \right|^2 }{2} \nonumber\\
 && + \frac{\gamma}{2} \left[ \frac{\left| \tilde{\mu}_2 \right|^2}{\tilde{\alpha}_1}
  \left( e^{\tilde{\alpha}_1(t-\Delta\tau)} - 1 \right)
  - \frac{ \tilde{\mu}_2 \tilde{\mu}_3^{*} }{\tilde{\alpha}_2} \left( e^{\tilde{\alpha}_2(t-\Delta\tau)} - 1 \right)
  - \frac{ \tilde{\mu}_3 \tilde{\mu}_2^{*} }{\tilde{\alpha}_3} \left( e^{\tilde{\alpha}_3(t-\Delta\tau)} - 1 \right)
  + \frac{\left| \tilde{\mu}_3 \right|^2}{\tilde{\alpha}_4} \left( e^{\tilde{\alpha}_4(t-\Delta\tau)} - 1 \right)  \right] \nonumber\\
 &&+ \left| \tilde{\mu}_1 \right|^2 \left| e^{\tilde{\omega}_{+}(t-\Delta\tau)} - e^{\tilde{\omega}_{-}(t-\Delta\tau)} \right|^2
  \frac{ \langle b^{\dagger}(\Delta\tau) b(\Delta\tau) \rangle + \langle b(\Delta\tau) b^{\dagger}(\Delta\tau) \rangle }{2} \nonumber\\
 && + \frac{\Gamma(1+2n_\text{th})}{2} \left| \tilde{\mu}_1 \right|^2
  \left[ \frac{e^{\tilde{\alpha}_1(t-\Delta\tau)} - 1}{\tilde{\alpha}_1}
  - \frac{e^{\tilde{\alpha}_2(t-\Delta\tau)} - 1}{\tilde{\alpha}_2}
  - \frac{e^{\tilde{\alpha}_3(t-\Delta\tau)} - 1}{\tilde{\alpha}_3}
  + \frac{e^{\tilde{\alpha}_4(t-\Delta\tau)} - 1}{\tilde{\alpha}_4}  \right].
\end{eqnarray}
The coefficients $\tilde{\alpha}_i$ can be given by
\begin{eqnarray}
\tilde{\alpha}_1 &=& \tilde{\omega}_{+} + \tilde{\omega}_{+}^{*}, \quad
\tilde{\alpha}_2 = \tilde{\omega}_{+} + \tilde{\omega}_{-}^{*}, \nonumber\\
\tilde{\alpha}_3 &=& \tilde{\omega}_{-} + \tilde{\omega}_{+}^{*}, \quad
\tilde{\alpha}_4 = \tilde{\omega}_{-} + \tilde{\omega}_{-}^{*}.
\end{eqnarray}
These correlation functions at time $\Delta\tau$ can be calculated as follows
\begin{eqnarray}
\langle a^{\dagger}(\Delta\tau)a(\Delta\tau) \rangle &=&
  (1+n_\text{th}) \left| \mu_1 \right|^2 \left| -e^{\omega_{+}\tau_1} + e^{\omega_{-}\tau_1} \right|^2
  e^{-\tilde{\gamma}\tau_d}  \nonumber\\
 &+& (1+n_\text{th})\Gamma \left| \mu_1 \right|^2 e^{-\tilde{\gamma}\tau_d}
  \left[ \frac{ e^{ \alpha_1 \tau_1 }-1 }{\alpha_1} - \frac{ e^{ \alpha_2 \tau_1 }-1 }{\alpha_2}
  - \frac{ e^{ \alpha_3 \tau_1 }-1 }{\alpha_3} + \frac{ e^{ \alpha_4 \tau_1 }-1 }{\alpha_4} \right], \nonumber\\
\langle b(\Delta\tau) b^{\dagger}(\Delta\tau) \rangle &=& (1+n_\text{th}) e^{-\Gamma\tau_d}
  \left| -\mu_3 e^{\omega_{+}\tau_1} + \mu_2 e^{\omega_{-}\tau_1} \right|^2
  + (1+n_\text{th})\left( 1 - e^{-\Gamma\tau_d} \right) \nonumber\\
 &+& \Gamma(1+n_\text{th}) e^{-\Gamma\tau_d} \left[
  \frac{\left| \mu_3 \right|^2}{\alpha_1} \left( e^{\alpha_1\tau_1} - 1 \right)
  - \frac{\mu_3\mu_2^{*}}{\alpha_2} \left( e^{\alpha_2\tau_1} - 1 \right)
  - \frac{\mu_2\mu_3^{*}}{\alpha_2} \left( e^{\alpha_3\tau_1} - 1 \right)
  + \frac{\left| \mu_2 \right|^2}{\alpha_1} \left( e^{\alpha_4\tau_1} - 1 \right)  \right], \nonumber\\
\langle a(\Delta\tau)b(\Delta\tau) \rangle &=&
  \left( \mu_2 e^{\omega_{+}\tau_1} - \mu_3 e^{\omega_{-}\tau_1} \right) \mu_1^{*}
  \left( e^{\omega_{+}^{*}\tau_1} - e^{\omega_{-}^{*}\tau_1} \right)
  e^{-(\tilde{\gamma}/2+i\Delta_a)\tau_d} e^{-(\Gamma/2+i\Delta_b)\tau_d} \nonumber\\
 &+& n_\text{th} \left( -\mu_3^{*} e^{\omega_{+}^{*}\tau_1} + \mu_2^{*} e^{\omega_{-}^{*}\tau_1} \right)
  \mu_1 \left( e^{\omega_{+}\tau_1} - e^{\omega_{-}\tau_1} \right)
  e^{-(\tilde{\gamma}/2+i\Delta_a)\tau_d} e^{-(\Gamma/2+i\Delta_b)\tau_d} \nonumber\\
 &+& \gamma\mu_1^{*} e^{-(\tilde{\gamma}/2+i\Delta_a)\tau_d} e^{-(\Gamma/2+i\Delta_b)\tau_d} \nonumber\\
 &\times&\left[ \frac{\mu_2(e^{\alpha_1\tau_1}-1)}{\alpha_1} - \frac{\mu_2(e^{\alpha_2\tau_1}-1)}{\alpha_2}
  - \frac{\mu_3(e^{\alpha_3\tau_1}-1)}{\alpha_3} +  \frac{\mu_3(e^{\alpha_4\tau_1}-1)}{\alpha_4} \right]  \nonumber\\
 &+& \Gamma n_\text{th} \mu_1 e^{-(\tilde{\gamma}/2+i\Delta_1)\tau_d} e^{-(\Gamma/2+i\Delta_2)\tau_d}  \nonumber\\
 &\times&\left[ \frac{\mu_3^{*}(e^{\alpha_1\tau_1}-1)}{\alpha_1} - \frac{\mu_2^{*}(e^{\alpha_2\tau_1}-1)}{\alpha_2}
  - \frac{\mu_3^{*}(e^{\alpha_3\tau_1}-1)}{\alpha_3} +  \frac{\mu_2^{*}(e^{\alpha_4\tau_1}-1)}{\alpha_4} \right], \nonumber\\
\langle a^{\dagger}(\Delta\tau) b^{\dagger}(\Delta\tau) \rangle &=&
  (1+n_\text{th}) \left( -e^{\omega_{+}^{*}\tau_1} + e^{\omega_{-}^{*}\tau_1} \right) \mu_1^{*}
  \left( -\mu_3 e^{\omega_{+}\tau_1} + \mu_2 e^{\omega_{-}\tau_1} \right)
  e^{-(\tilde{\gamma}/2-i\Delta_a)\tau_d} e^{-(\Gamma/2-i\Delta_b)\tau_d} \nonumber\\
 &+& \Gamma(1+n_\text{th})\mu_1^{*} e^{-(\tilde{\gamma}/2-i\Delta_a)\tau_d} e^{-(\Gamma/2-i\Delta_b)\tau_d} \nonumber\\
 &\times& \left[ \frac{\mu_3(e^{\alpha_1\tau_1}-1)}{\alpha_1}
  - \frac{\mu_3(e^{\alpha_2\tau_1}-1)}{\alpha_2}
  - \frac{\mu_2(e^{\alpha_3\tau_1}-1)}{\alpha_3}
  + \frac{\mu_2(e^{\alpha_4\tau_1}-1)}{\alpha_4} \right].
\end{eqnarray}
Therefore, the logarithmic negativity $\tilde{E}_{\mathcal{N}}$ can be calculated as follows
\begin{eqnarray}
\tilde{E}_{\mathcal{N}} = {\rm max} \left[ 0,\ {\rm ln}2\tilde{\lambda}_{-} \right],
\end{eqnarray}
where
\begin{eqnarray}
\tilde{\lambda}_{-} = \sqrt{ \frac{ \Sigma(\tilde{\mathcal{V}})
- \sqrt{ \left( \Sigma(\tilde{\mathcal{V}}) \right)^2 - 4\det\tilde{\mathcal{V}} } }{2} },
\end{eqnarray}
and
\begin{eqnarray}
\Sigma(\tilde{\mathcal{V}}) &=& \tilde{\mathcal{V}}_{11}^2 + \tilde{\mathcal{V}}_{33}^2
+ 2\left( \tilde{\mathcal{V}}_{13}^2 + \tilde{\mathcal{V}}_{14}^2 \right), \nonumber\\
\det\tilde{\mathcal{V}} &=& \left( \tilde{\mathcal{V}}_{13}^2 + \tilde{\mathcal{V}}_{14}^2
- \tilde{\mathcal{V}}_{11} \tilde{\mathcal{V}}_{33} \right)^2.
\end{eqnarray}
As we consider the system parameters in the regime that $g\gg\Gamma\gg\gamma$,
$\Gamma\gg\Delta_a=\Delta_{\tilde{a}}\gg\Delta_b$, $\tau_{d}\gg\tau_{1,2}$, and environmental temperature is
tens of Kelvin, after some simplifications we have
\begin{eqnarray}\label{Simplified En2}
\tilde{\lambda}_{-} &\approx& \left| \frac{ \left( \tilde{\mathcal{V}}_{11} \tilde{\mathcal{V}}_{ss} - \tilde{\mathcal{V}}_{33}^2 \right)
\left( \tilde{\mathcal{V}}_{11} + \tilde{\mathcal{V}}_{33} \right) }
{ \tilde{\mathcal{V}}_{11}^2 + \tilde{\mathcal{V}}_{33}^2 + \tilde{\mathcal{V}}_{11} \tilde{\mathcal{V}}_{33}
+ \tilde{\mathcal{V}}_{13}^2 } \right|,
\end{eqnarray}
where
\begin{eqnarray}\label{Simplified V2}
\tilde{\mathcal{V}}_{11} &\approx& \frac{1+2n_s}{2}, \nonumber\\
\tilde{\mathcal{V}}_{33} &\approx& \frac{1+2n_\text{th}}{2} + \left[ \frac{n_{b}-2n_\text{th}}{2}
  - \frac{n_b}{2}\cos\left(2\tilde{g}(t-\Delta\tau )\right)
  - \frac{\Gamma(1+2n_\text{th})}{8\tilde{g}} \sin\left(2\tilde{g}(t-\Delta\tau )\right) \right] e^{-\frac{\Gamma}{2}(t-\Delta\tau)}, \nonumber\\
\tilde{\mathcal{V}}_{13} &\approx& \frac{\mathcal{C}_{ns}}{2} \sin\left(\tilde{g}(t-\Delta\tau )\right) e^{-\frac{\Gamma}{4}(t-\Delta\tau)},
\end{eqnarray}
and
\begin{eqnarray}
n_s &=& \langle a^{\dagger}(\Delta\tau)a(\Delta\tau) \rangle, \nonumber\\
n_b &=& \langle b^{\dagger}(\Delta\tau)b(\Delta\tau) \rangle, \nonumber\\
\mathcal{C}_{ns} &=& {\rm Im}\left[ \langle a(\Delta\tau)b(\Delta\tau) \rangle
+ \langle b(\Delta\tau) a(\Delta\tau) \rangle  \right].
\end{eqnarray}
Actually, $n_s$ and $n_b$ denote the Stokes photon occupation and
acoustic phonon occupation at time $\tau_1+\tau_d$. $\mathcal{C}_{ns}$ corresponds to the cross-correlation
between Stokes photons and acoustic phonons at time $\tau_1+\tau_d$.
Substituting Eqs.~(\ref{Simplified V2}) into Eq.~(\ref{Simplified En2}), $\tilde{\lambda}_{-}$
can be simplified as follows
\begin{eqnarray}\label{Simplified En2 with higher-order terms}
\tilde{\lambda}_{-} &\approx& \frac{1}{2}\left| \frac{ \sum\limits_{n=0}^{12} \eta_{1n}(t-\Delta\tau )^{n} }
  { \sum\limits_{n=0}^{14} \eta_{2n}(t-\Delta\tau )^{n} } \right| ,
\end{eqnarray}
with
\begin{eqnarray}\label{Simplified eta1 with higher-order terms}
\eta_{10} &=& \frac{ 1 + 3n_s + 2n_s^2 }{4}, \nonumber\\
\eta_{11} &=& -\frac{\Gamma}{16} ( 3 + 8n_s + 4n_s^2 ), \nonumber\\
\eta_{12} &=& \tilde{g}^2\frac{ n_{b}( 3 + 8n_s + 4n_s^2 ) - (1+n_s)\mathcal{C}_{ns}^2 }{4}, \nonumber\\
\eta_{13} &=& \Gamma\tilde{g}^2\frac{ (3+2n_s)( 2 + 4n_s + 3\mathcal{C}_{ns}^2 )
  + 4n_\text{th}( 3 + 8n_s + 4n_s^2 ) - 6 n_b( 5 + 12n_s + 4n_s^2 ) }{48}, \nonumber\\
\eta_{14} &=& \tilde{g}^4\frac{ 6n_b^2(1+2n_s) - n_b( 3 + 8n_s + 4n_s^2 )
  + ( 1 + n_s - 3n_b )\mathcal{C}_{ns}^2 }{12}, \nonumber\\
\eta_{15} &=& -\Gamma\tilde{g}^4(1+2n_s) \frac{ 3 + 2n_s
  + 5n_b( 12n_b - 2n_s - 9 ) + n_\text{th}( 6 - 40n_b + 4n_s ) }{120} \nonumber\\
 && - \Gamma\tilde{g}^4 \frac{ 5 + 4n_\text{th} - 12n_b + 2n_s }{48} \mathcal{C}_{ns}^2, \nonumber\\
\eta_{16} &=& \tilde{g}^6 \frac{ n_b( 3 + 8n_s + 4n_s^2 )
  - 30n_b^2( 1+2n_s ) - ( 1 + n_s - 15n_b )\mathcal{C}_{ns}^2 }{90}, \nonumber\\
\eta_{17} &=& \Gamma\tilde{g}^2(1+2n_s) \frac{ 3 + 420n_b^2 + 2n_s
  - 7n_b( 21+2n_s ) + 2n_\text{th}( 3 - 112n_b + 2n_s ) }{1260} \nonumber\\
 && + \Gamma\tilde{g}^6 \frac{ 11 + 16n_\text{th} - 60n_b + 2n_s }{360} \mathcal{C}_{ns}^2, \nonumber\\
\eta_{18} &=& \tilde{g}^8 \frac{ 126n_b^2(1+2n_s)
  - n_b( 3 + 8n_s + 4n_s^2 ) + ( 1 + n_s - 63n_b )\mathcal{C}_{ns}^2 }{120}, \nonumber\\
\eta_{19} &=& -\Gamma\tilde{g}^8(1+2n_s) \frac{ 3 + 2n_s
  + 3n_b( 756n_b - 6n_s - 179 ) + 2n_\text{th}( 3 - 492n_b + 2n_s ) }{22680} \nonumber\\
 && - \Gamma\tilde{g}^8 \frac{ 91 + 164n_\text{th} - 756n_b + 6n_s }{15120} \mathcal{C}_{ns}^2, \nonumber\\
\eta_{110} &=& \tilde{g}^{10} \frac{ n_b( 3 + 8n_s + 4n_s^2 )
  - 510n_b^2( 1+2n_s ) - ( 1 + n_s - 256n_b )\mathcal{C}_{ns}^2 }{28350}, \nonumber\\
\eta_{111} &=& \Gamma\tilde{g}^{10} \frac{ n_b(1+2n_s)( -3 + 4n_\text{th} + 1020n_b - 2n_s )
  + ( 1 - n_\text{th} - 510n_b + n_s )S^2 }{56700}, \nonumber\\
\eta_{112} &=& \tilde{g}^{12} \frac{ n_b(1+2n_s)(2n_\text{th}+1023n_b) }{467775}
  - \tilde{g}^{12} \frac{(n_\text{th}+1023n_b) \mathcal{C}_{ns}^2}{935550},
\end{eqnarray}
and
\begin{eqnarray}
\eta_{20} &=& \frac{ 1 + 3n_s + 2n_s^2 }{2}, \nonumber\\
\eta_{21} &=& -\frac{\Gamma(3+2n_s)}{8}, \nonumber\\
\eta_{22} &=& \tilde{g}^2 \frac{ 2n_b (3+2n_s) + \mathcal{C}_{ns}^2  }{4}, \nonumber\\
\eta_{23} &=& \Gamma\tilde{g}^2 \frac{ 6 + 4n_s + 4n_\text{th}(3+2n_s)
  - 6n_b(5+2n_s) - 3\mathcal{C}_{ns}^2 }{24}, \nonumber\\
\eta_{24} &=& \tilde{g}^4 \frac{ 12n_b^2 - 2n_b(3+2n_s) - \mathcal{C}_{ns}^2 }{12}, \nonumber\\
\eta_{25} &=& \Gamma\tilde{g}^4 \frac{ -3 - 2n_s
  - 5n_b( 12n_b - 9 - 2n_s ) + n_\text{th}( -6 + 40n_b - 4n_s ) }{60}, \nonumber\\
\eta_{26} &=& \tilde{g}^6 \frac{ -60n_b^2 + 2n_b( 3+2n_s ) + \mathcal{C}_{ns}^2 }{90}, \nonumber\\
\eta_{27} &=& \Gamma\tilde{g}^6 \frac{ 6 + 4n_s
  + 14n_b( 60n_b - 21 - 2n_s ) + 4n_\text{th}( 3 - 112n_b + 2n_s ) - 7\mathcal{C}_{ns}^2 }{1260}, \nonumber
\end{eqnarray}
\begin{eqnarray}\label{Simplified eta1 with higher-order terms}
\eta_{28} &=& \tilde{g}^8 \frac{ 252n_b - 2n_b(3+2n_s) - \mathcal{C}_{ns}^2 }{1260}, \nonumber\\
\eta_{29} &=& \Gamma\tilde{g}^8 \frac{ -6 - 4n_s
  - 6n_b( 756n_b - 6n_s - 179 ) + 4n_\text{th}( -3 + 492n_b - 2n_s ) + 9\mathcal{C}_{ns}^2 }{22680},  \nonumber\\
\eta_{210} &=& \tilde{g}^{10} \frac{ 2n_b( 3+2n_s ) - 1020n_b^2 + \mathcal{C}_{ns}^2 }{28350}, \nonumber\\
\eta_{211} &=& -\Gamma\tilde{g}^{10} \frac{ 2n_b( 3 - 4n_\text{th} + 2n_s ) - 2040n_b^2 + \mathcal{C}_{ns}^2 }{56700}, \nonumber\\
\eta_{212} &=& \tilde{g}^{12} \frac{ n_b(2n_\text{th}+1023n_b) }{233888}, \nonumber\\
\eta_{213} &=& -\Gamma\tilde{g}^{12} \frac{ n_b (2n_\text{th}+1023n_b) }{233888}, \nonumber\\
\eta_{214} &=& - \tilde{g}^{14}\frac{n_b^2}{2598}.
\end{eqnarray}

\begin{figure}[h]
\centerline{
\includegraphics[width=9 cm,clip]{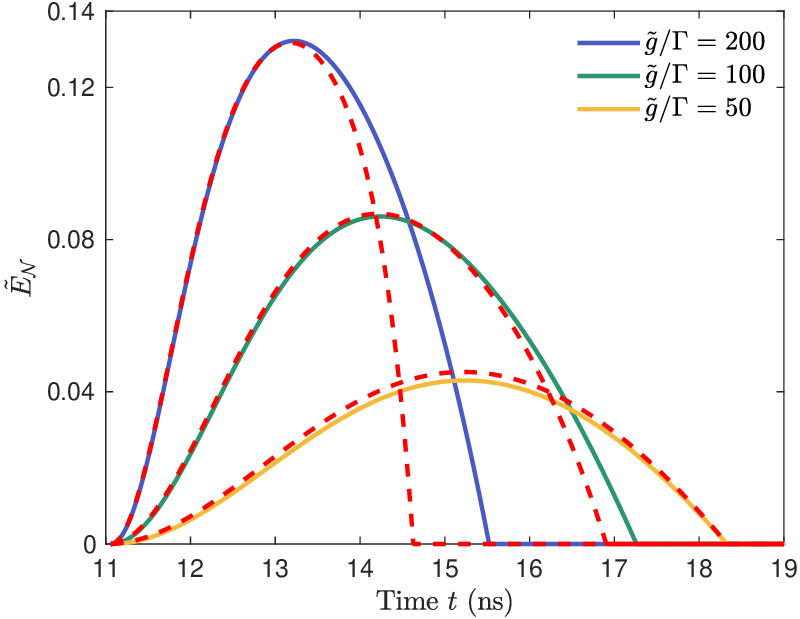}}
\caption{(Color online) Time evolution of $\tilde{E}_{\mathcal{N}}$ for different coupling strength at temperature $30~$K,
where solid curves and dashed curves denote the accurate simulation and corresponding approximated results, respectively.
}\label{Sup_Fig4}
\end{figure}

We present the time evolution of $\tilde{E}_{\mathcal{N}}$ in Fig.~\ref{Sup_Fig4}, where the solid curves denote
the accurate solutions of $\tilde{E}_{\mathcal{N}}$ versus different coupling strength $\tilde{g}$ and red dashed
curves correspond to the approximated solutions evaluated by Eq.~(\ref{Simplified En2 with higher-order terms}).
Here, we assume that the coupling strength during the entanglement generation is $g/\Gamma=30$, temperature is $T_\text{m}=30~$K,
and delayed time is $\tau_d=0.1~$ns.

\begin{figure}[h]
\centerline{
\includegraphics[width=9 cm,clip]{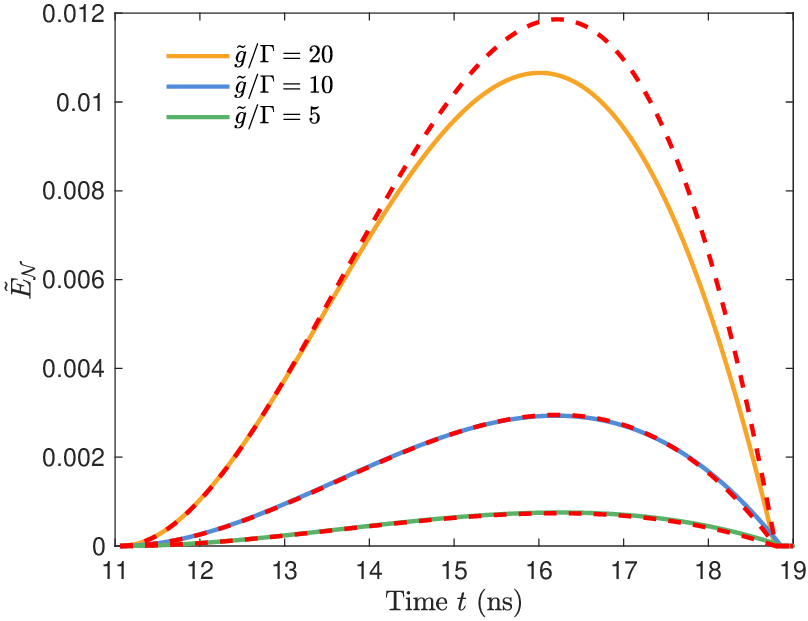}}
\caption{(Color online) Time evolution of $\tilde{E}_{\mathcal{N}}$ for different coupling strength at temperature $30~$K,
where solid curves denote the accurate simulation results and red dashed curves correspond to the approximated analytical
results evaluated by Eq.~(\ref{Final simplified En2}).
}\label{Sup_Fig5}
\end{figure}

However, the approximated expression of Eq.~(\ref{Simplified En2 with higher-order terms}) is still complicated.
If we consider the small time $\tilde{g}(t-\Delta\tau )\ll1$, Eq.~(\ref{Simplified En2 with higher-order terms})
can be further simplified by ignoring higher-order terms as follows
\begin{eqnarray}\label{Simplified En2 without higher-order terms}
\tilde{\lambda}_{-} \approx \frac{1}{2} \left| \frac{ 1 + \tilde{\eta}_1 (t-\Delta\tau )^2 + \tilde{\eta}_2^2 (t-\Delta\tau )^3 }
{ 1 + \tilde{\eta}_3 (t-\Delta\tau )^2 } \right|,
\end{eqnarray}
with
\begin{eqnarray}
\tilde{\eta}_1 &=& \tilde{g}^2 \left[ \frac{n_b}{1+n_s} + \frac{2n_b(1+2n_s) - \mathcal{C}_{ns}^2}{1+2n_s} \right], \nonumber\\
\tilde{\eta}_2 &=& \frac{  2\tilde{g}^2(\Gamma n_\text{th}) }{3}, \nonumber\\
\tilde{\eta}_3 &=& \tilde{g}^2 \left[ \frac{n_b}{1+n_s} + \frac{4n_b + \mathcal{C}_{ns}^2}{2(1+n_s)(1+2n_s)} \right] .
\end{eqnarray}
If $\sqrt{\tilde{\eta}_3}(t-\Delta\tau)\ll1$, Eq.~(\ref{Simplified En2 without higher-order terms}) can be
further reduced to
\begin{eqnarray}\label{Final simplified En2}
\tilde{\lambda}_{-} \approx \frac{1}{2} \left[ 1 - \eta^2 (t-\Delta\tau)^2 + \frac{2}{3}\tilde{g}^2(\Gamma n_\text{th})(t-\Delta\tau)^3 \right],
\end{eqnarray}
where
\begin{eqnarray}
\eta^2 &=& \frac{\tilde{g}^2(\mathcal{C}_{ns}^2-4n_{b}n_{s})}{1+2n_{s}}.
\end{eqnarray}
We present the accurate simulation results and approximated analytical results evaluated by Eq.~(\ref{Final simplified En2}) in Fig.~\ref{Sup_Fig5}
In the main text, we assume that the initial time of the entanglement readout process is
at $\tau_1+\tau_d$, i.e., $t\in [0,\tau_1+\tau_d]$, thus Eq.~(\ref{Final simplified En2}) can be rewritten as follows
\begin{eqnarray}
\tilde{\lambda}_{-} \approx \frac{1}{2} \left[ 1 - \eta^2 t^2 + \frac{2}{3}\tilde{g}^2(\Gamma n_\text{th})t^3 \right].
\end{eqnarray}

\end{document}